\documentclass[twocolumn, twocolappendix]{aastex631}

\usepackage{bm}
\usepackage{amsmath}
\usepackage{booktabs}

\newcommand{\nh}{n_{\text{H}}} %

\newcommand{\pcmq}{{\rm cm^{-3}}}
\newcommand{\psec}{{\rm s^{-1}}}
\newcommand{\erg}{{\rm erg}}
\newcommand{\kelvin}{{\rm K}}
\newcommand{\Msun}{{\rm  M_{\odot}}}
\newcommand{\kB}{{k_{\rm B}}}

\newcommand{\Zsun}{Z_{\odot}}

\newcommand{\hikpc}{{h^{-1}\,{\rm kpc}}}
\newcommand{\hicmpc}{{h^{-1}\,{\rm cMpc}}}
\newcommand{\hickpc}{{h^{-1}\,{\rm ckpc}}}
\newcommand{\himsun}{{h^{-1}\,{\Msun}}}

\newcommand{\kms}{{\rm km\,s^{-1}}}
\newcommand{\osakasim}{CROCODILE} %

\shorttitle{Osaka Feedback Model III: Cosmological Run}
\shortauthors{Oku \& Nagamine}

\graphicspath{{./}{figures/}}

\begin{document}

\title{Osaka Feedback Model III: Cosmological Simulation CROCODILE}

\correspondingauthor{Yuri Oku}
\email{oku@astro-osaka.jp}

\author[0000-0002-5712-6865]{Yuri Oku}
\affiliation{Theoretical Astrophysics, Department of Earth and Space Science, Graduate School of Science, Osaka University, \\
1-1 Machikaneyama, Toyonaka, Osaka 560-0043, Japan}
\affiliation{Center for Cosmology and Computational Astrophysics, the Institute for Advanced Study in Physics,  Zhejiang University, China} 
\author[0000-0001-7457-8487]{Kentaro Nagamine}
\affiliation{Theoretical Astrophysics, Department of Earth and Space Science, Graduate School of Science, Osaka University, \\
1-1 Machikaneyama, Toyonaka, Osaka 560-0043, Japan}
\affiliation{Theoretical Joint Research, Forefront Research Center, Graduate School of Science, Osaka University, Toyonaka, Osaka 560-0043, Japan}
\affiliation{Kavli IPMU (WPI), The University of Tokyo, 5-1-5 Kashiwanoha, Kashiwa, Chiba 277-8583, Japan}
\affiliation{Department of Physics \& Astronomy, University of Nevada, Las Vegas, 4505 S. Maryland Pkwy, Las Vegas, NV 89154-4002, USA}
\affiliation{Nevada Center for Astrophysics, University of Nevada, Las Vegas, 4505 S. Maryland Pkwy, Las Vegas, NV 89154-4002, USA}

\begin{abstract}

We introduce our new cosmological simulation dataset {\osakasim}, executed using the \textsc{GADGET4-Osaka}  smoothed particle hydrodynamics code. This simulation incorporates an updated supernova (SN) feedback model of  \citet{2022ApJS..262....9O} and an active galactic nuclei (AGN) feedback model. 
{A key innovation in our SN feedback model is the integration of a metallicity- and redshift-dependent, top-heavy IMF. Our SN model introduces a new consideration that results in an order of magnitude difference in the energy injection rate per unit stellar mass formed at high redshift.}
The {\osakasim} dataset is comprehensive, encompassing a variety of runs with diverse feedback parameters. This allows for an in-depth exploration of the relative impacts of different feedback processes in galactic evolution.
Our initial comparisons with observational data, spanning the galaxy stellar mass function, the star formation main sequence, and the mass-metallicity relation, show promising agreement, especially for the Fiducial run. These results establish a solid foundation for our future work. 
We find that SN feedback is a key driver in the chemical enrichment of the IGM.
Additionally, the AGN feedback creates metal-rich, bipolar outflows that extend and enrich the CGM and IGM over a few Mpc scales.

\end{abstract}


\section{Introduction}

The theory of structure formation in the Universe has been extensively developed within the framework of the $\Lambda$ cold dark matter ($\Lambda$CDM) model. This model provides robust predictions regarding the distribution of matter on large scales, where the gravitational influence of dark matter is the predominant force driving these distributions \citep[e.g.,][]{1984Natur.311..517B,1985ApJ...292..371D,1995Natur.377..600O, 1998Natur.391...51P,1999Sci...284.1481B}. 

However, this model encounters intricacies at smaller scales. Here, the effects of astrophysical and hydrodynamical processes become increasingly significant, 
adding complexity to our understanding of baryonic matter distribution and the evolution and formation of galaxies and black holes throughout cosmic history.

To address these complexities, cosmological hydrodynamical simulations have emerged as a powerful and direct approach. These simulations integrate gravity and hydrodynamics within $\Lambda$CDM cosmology, while accounting for a range of astrophysical effects. This methodology has proved instrumental in advancing our comprehensive understanding of structure formation \citep[for a technical review, see][]{2020NatRP...2...42V}.

Recent advancements in cosmological hydrodynamical simulations have been successful in replicating key observed galaxy statistics, such as the galaxy stellar mass function. 
Pioneering simulations like Illustris \citep{2014MNRAS.445..175G}, Magneticum \citep{2014MNRAS.442.2304H}, EAGLE \citep{2015MNRAS.446..521S}, MassiveBlack-II \citep{2015MNRAS.450.1349K}, Horizon-AGN \citep{2017MNRAS.467.4739K}, MUFASA \citep{2016MNRAS.462.3265D}, Romulus \citep{2017MNRAS.470.1121T}, IllustrisTNG \citep{2018MNRAS.475..648P}, SIMBA \citep{2019MNRAS.486.2827D}, ASTRID \citep{2022MNRAS.512.3703B}, FIREbox \citep{2023MNRAS.522.3831F}, MillenniumTNG \citep{2023MNRAS.524.2539P}, and FLAMINGO \citep{2023MNRAS.526.4978S} have demonstrated considerable agreement with the observed galaxy stellar mass functions across various redshifts, although achieving a perfect match at every redshift remains an ongoing challenge, reflecting the complexity and dynamic nature of galaxy evolution. 

These simulations assume concordance $\Lambda$ CDM cosmology and incorporate an array of sophisticated subgrid models, including radiative cooling, extragalactic ultraviolet (UV) background radiation, star formation, and supernova (SN) feedback. 
Furthermore, several of these simulations also integrate active galactic nuclei (AGN) feedback, adding another layer of complexity and realism to the simulation of galactic behaviors and evolution.

The diverse yet consistent successes of various independent simulation projects, each employing distinct subgrid models, have fostered a broad consensus within the scientific community.
This consensus underscores the critical role of supernova (SN) and active galactic nuclei (AGN) feedback in galaxy formation.
The ability of these simulations to accurately reproduce the observed galaxy stellar mass functions, particularly through the calibration of feedback models, further highlights the pivotal role these phenomena play in galaxy formation \citep{2015ARA&A..53...51S, 2017ARA&A..55...59N}.

However, there are still discrepancies in the properties of CGM and IGM among simulations and observations.
\citet{2023MNRAS.519.2251B} have used the dataset from the CAMELS project \citep{2023ApJS..265...54V} to study the \textsc{O\,vii} column density at low redshift ($z<0.3$) in simulations performed with the same code as used for IllustrisTNG and SIMBA simulations, and compared them with the stacked \textit{Chandra} observations of X-ray absorption lines on a sightline of a background quasar by \citet{2019ApJ...872...83K}.
They found that SIMBA predicts a lower \textsc{O\,vii} column density for a given \textsc{H i} column density than the IllustrisTNG model by an order of magnitude.
They also found that the \textsc{O\,vii} column density in simulations is lower than the observed one by \citet{2019ApJ...872...83K} by more than an order of magnitude, even for all ranges of SN and AGN feedback parameters explored in CAMELS.
\citet{2024ApJ...962...29S} analyzed a suite of zoom-in simulations conducted as part of the AGORA project \citep{2021ApJ...917...64R}. These simulations were performed using various simulation codes, both grid-based and particle-based codes, with different feedback models calibrated to match the semi-empirical models' predicted stellar-to-halo mass ratio values. 
They investigated the column densities of four ions, namely Si\textsc{\,iv}, \textsc{C\,iv}, \textsc{O\,vi}, and Ne\,\textsc{viii}, in the CGM.
They found that there were significant variations in the ion column densities between the simulations and the observations. At $z=3$, the simulations showed a difference of three orders of magnitude compared to the observations.
They also found that the column densities from grid-based code tend to be higher than those from particle-based code, with varying scatter due to different feedback prescriptions.
The difference in the properties of the CGM and IGM found in these works arises from the feedback models employed in the simulations. Therefore, a better modeling of feedback based on small-scale physics is necessary. 

In \citet{2022ApJS..262....9O}, we have developed an SN feedback model based on high-resolution simulations. 
In cosmological simulations, the adiabatic phase of an SN remnant is rarely resolved, and then the hydro solver cannot solve the conversion between the kinetic and thermal energy.
Thus, we consider the effect of SN kinetic and thermal feedback on the ISM and CGM/IGM scale, respectively \citep[see also][]{2023MNRAS.523.3709C}.
The kinetic feedback considers the momentum of a superbubble formed by clustered SNe, which drives interstellar turbulence.
The thermal feedback considers the fact that some superbubbles break out from the galactic disk to generate hot galactic wind as observed, e.g., M82 \citep{2009ApJ...697.2030S}.
In Milky Way-mass isolated galaxy simulations, we have demonstrated that the kinetic feedback supports a galactic disk against the gravitational collapse to suppress star formation, and the thermal feedback drives hot metal-rich galactic wind.

In this paper, we introduce the {\osakasim}\footnote{Named in homage to Osaka University's official mascot, Dr. Wani, a crocodile (\url{https://www.osaka-u.ac.jp/sp/drwani/en/})}\footnote{The project's webpage is \url{https://sites.google.com/view/crocodilesimulation/home}} (Cosmological hydROdynamical simulation of struCture fOrmation and feeDback physIcs in gaLaxy Evolution) cosmological simulation. 
This simulation, conducted using the \textsc{GADGET4-Osaka} smoothed particle hydrodynamics (SPH) code \citep[][a modified version of \textsc{GADGET-4},  \citealt{2021MNRAS.506.2871S}]{2022MNRAS.514.1441R, 2022MNRAS.514.1461R}, features our updated SN feedback model from \citet{2022ApJS..262....9O} and incorporates the active galactic nuclei (AGN) feedback model following  \citet{2015MNRAS.454.1038R, 2015MNRAS.446..521S, 2015MNRAS.450.1937C}. 
The primary aims of this paper are twofold: firstly, to demonstrate {\osakasim}'s capability in accurately reproducing essential galaxy statistics; and secondly, to elucidate the impacts of SN and AGN feedback on the metal enrichment of the IGM.

The study of metal enrichment in the IGM offers a distinct and robust means of constraining feedback physics. 
This is because the spatial distribution and abundance patterns of metals in the IGM serve as historical records of feedback activities. 
The IGM is dilute and difficult to observe directly. 
However, it is possible to reconstruct the three-dimensional distribution of foreground absorbers from absorption lines on multiple lines of sight of galaxies and quasars, which is known as `IGM tomography' \citep{2014ApJ...795L..12L, 2018ApJS..237...31L, 2020ApJ...891..147N}.
The advent of future wide-field and high-spectral-resolution IGM tomography surveys promises to revolutionize our understanding. 
These surveys will utilize advanced multiplexed fiber spectrographs such as Subaru/PFS \citep{2022arXiv220614908G}, WHT/WEAVE \citep{2023MNRAS.tmp..715J}, VLT/MOONS \citep{2014SPIE.9147E..0NC}, MSE \citep{2019arXiv190404907T}, ELT/ANDES \citep{2013arXiv1310.3163M} and ELT/MOSAIC \citep{2019A&A...632A..94J}.
These innovative tools are expected to unveil the intricate three-dimensional metal distribution within the IGM, offering unprecedented insights into cosmic structure formation.

The prediction of the matter distribution from numerical simulations is indispensable to deduce information on feedback physics from IGM tomographic observations. 
The potential of IGM tomography as a powerful tool for probing feedback is highlighted by \citet{2021ApJ...914...66N}.
Using \textsc{GADGET3-Osaka} cosmological simulation \citep{2019MNRAS.484.2632S}, their study revealed variations in the Ly$\alpha$ optical depth on small scales. 
These variations were influenced by the specifics of the feedback model, as well as the treatment of gas self-shielding, star formation, and ultraviolet background radiation (UVB).
While \citet{2021ApJ...914...66N} primarily focused on the distribution of neutral hydrogen, examining the metal distribution can provide further insight into the feedback physics.
As a first step, in this paper, we investigate the overall metal distribution in the IGM by analyzing its power spectrum. A more detailed analysis of individual ion species is planned for our future studies.

The remainder of this article is organized as follows: 
Section~\ref{sec:method} details our simulation methodology, including the subgrid models for stellar feedback and black holes.
In Section~\ref{sec:galaxystatistics}, we present the results of our simulations, with a focus on the basic statistical properties of galaxies to validate our simulation framework.
In Section~\ref{sec:IGM}, we analyze the distribution of metals in the IGM, highlighting the impact of SN and AGN feedback. This section serves as a preliminary study, setting the theoretical groundwork for future tomographic surveys.
Our findings are summarized and future research directions are outlined in Section~\ref{sec:summary}.
Additionally, Appendix~\ref{sec:hydrodynamics} discusses the numerical methodology of our hydrodynamical simulation. 

\section{Method}
\label{sec:method}
In this section, we describe the methodology of the simulation.

\subsection{Initial Condition}
We generate an initial condition using \textsc{MUSIC2-monofonIC}\footnote{The code's website is \url{https://bitbucket.org/ohahn/monofonic/src/master/}} \citep{2021MNRAS.500..663M, 2021MNRAS.503..426H}.
The phase fixing technique \citep{2016MNRAS.462L...1A} is used to suppress the impact of cosmic variance, but we run only a single simulation for each model and do not include paired simulations for the analyses in this paper.
We adopt the following cosmological parameters from  \citet{2020A&A...641A...6P}, ($\Omega_m$, $\Omega_b$, $\Omega_\Lambda$, $h$, $n_s$, $10^9A_s$) = (0.3099, 0.0488911, 0.6901, 0.67742, 0.96822, 2.1064).
\footnote{The best-fit parameters of the baseline model 2.20 base\_plikHM\_TTTEEE\_lowl\_lowE\_lensing\_post\_BAO\_Pantheon of Planck 2018 cosmological parameter table as of May 14, 2019 (\url{https://wiki.cosmos.esa.int/planck-legacy-archive/images/b/be/Baseline_params_table_2018_68pc.pdf}).}
The initial condition is generated at $z_{\rm ini}=39$ using the third-order Lagrangian perturbation theory\,(3LPT).
The transfer function at $z_{\rm ini}$ is obtained by back-scaling the transfer function at the reference redshift $z_{\rm ref}=2$ computed by \textsc{CLASS}
\footnote{The code's website is \url{http://class-code.net/}} \citep{2011JCAP...07..034B}.
The simulation box size is $L_{\rm box}=50\,\hicmpc$, and the total number of particles is $N_{\rm p}=2\times512^3$; the initial particle mass of dark matter and gas is $m_{\rm DM}=6.74\times10^7\,\himsun$ and $m_{\rm gas}=1.26\times10^7\,\himsun$, respectively.

\subsection{Cosmological Hydrodynamics Simulation}
We use the cosmological N-body/SPH code \textsc{GADGET4-Osaka} \citep{2022MNRAS.514.1441R, 2022MNRAS.514.1461R}, which is a privately developed branch of \textsc{GADGET-4}\footnote{The code's website is \url{https://wwwmpa.mpa-garching.mpg.de/gadget4/}} \citep{2021MNRAS.506.2871S}.
The Newtonian gravity for dark matter and baryons is solved 
using the third-order TreePM method \citep{1995ApJS...98..355X, 2002JApA...23..185B, 2005MNRAS.364.1105S}. 
The same gravitational softening length is used for all particle types, and it is set to $\epsilon_{\rm grav}=3.38\,\hickpc$ but limited to physical $0.5\,\hikpc$ at all times.

The SPH method \citep[for reviews, see][]{2009NewAR..53...78R, 2010ARA&A..48..391S, 2012JCoPh.231..759P} is used to solve the governing equations of hydrodynamics.
Appendix \ref{sec:hydrodynamics} describes the numerical treatment of hydrodynamics.

The heating rate, $\Gamma$, and the cooling rate, $\Lambda$, are computed using the \textsc{Grackle}\footnote{The code's website is \url{https://grackle.readthedocs.io/en/latest/}} library \citep{2017MNRAS.466.2217S}, which solves the non-equilibrium primordial chemistry network of 12 chemical species, H, D, He, H$_2$, HD, and their ions, with radiative cooling by metals and photo-heating and photo-ionization by uniform ultraviolet background (UVB).
The metal cooling rate is precomputed using the photoionization code \textsc{Cloudy} \citep{2013RMxAA..49..137F}, and the UVB model by \citet{2012ApJ...746..125H} is assumed.
In \textsc{Grackle}, we set the parameters related to the onset of UVB to \texttt{UVbackground\_redshift\_on=8} and \texttt{UVbackground\_redshift\_fullon=6}, and then the cosmic reionization occurs at $z\sim7$. 
Our simulations do not include the heating by the cosmic rays and stellar radiation, and we ignore the self-shielding of the UVB by atomic hydrogen
to avoid overcooling of the neutral gas. 

We follow 12 metal elements, H, He, C, N, O, Ne, Mg, Si, S, Ca, Fe, and Ni, produced by core-collapse SNe, type-Ia SNe, and AGB stars.
The metallicity floor is $Z_{\rm floor} = 10^{-6}\,\Zsun$, where $\Zsun = 0.0134$ is the solar metallicity \citep{2009ARA&A..47..481A}.

\textsc{GADGET4-Osaka} solves the production and destruction of dust as described in \citet{2022MNRAS.514.1441R} following the full dust grain size distribution; however, we do not focus on dust in this work and omit the description of the dust module here. 
The analysis of dust will be made in our future work.

We use the Smagorinsky-Lilly type turbulent metal and dust diffusion model \citep{smagorinsky1963, 2010MNRAS.407.1581S, 2022MNRAS.514.1441R} with a diffusion parameter $C_{\rm diff} = 2\times10^{-4}$.

The nonthermal Jeans pressure floor \citep{2011MNRAS.417..950H, 2016ApJ...833..202K} is applied to avoid artificial fragmentation as described in the Appendix \ref{sec:jeans_floor}.
The lower limit of the temperature is set to the higher value of the CMB temperature and $T_{\rm min}=15$\,K, and the upper limit is set to $T_{\rm max}=10^{9}$\,K.

The simulations presented in this article are performed at the SQUID supercomputer at Cybermedia Center, Osaka University, equipped with dual Intel Xeon Platinum 8368 processors (2.4\,GHz) per compute node. Running the Fiducial model took $1.12\times10^5$ CPU hours.

In the following subsections, we describe our treatment of subgrid star formation, stellar feedback, and black hole physics. In this work, SNe and AGB stars are considered as sources of stellar feedback, while early stellar feedback, e.g., stellar radiation and stellar wind, is neglected. The adopted values of parameters introduced in the following subsections are summarized in Table \ref{tab:parameters}.

\begin{table*}
\centering
\caption{List of common parameters for subgrid physics}
\begin{tabular}{ccl}
\hline
Parameter & Adopted value & Description\\
\hline
$n_{\rm thres}$ & $0.1\,\pcmq$ & Lower density threshold to allow star formation\\
$T_{\rm thres}$ & $10^4\,\kelvin$ & Upper temperature threshold to allow star formation\\
$\epsilon_*$ & 0.01 & Star formation efficiency\\
$n_{\rm spawn}$ & 2 & Maximum number of stars spawned from one gas particle\\
$N_{\rm ngb, fb}$ & 8$\pm$2 & Number of gas particles subject to feedback\\
$n_{\rm event, snii}$ & 2 & SNII feedback event number\\
$n_{\rm event, snia}$ & 8 & SNIa feedback event number\\
$n_{\rm event, agb}$ & 8 & AGB feedback event number\\
$M_{\rm seeding, FoF}$  & $10^{10}\,\himsun$ & FoF mass threshold to seed BH \\
$M_{\rm seeding, star}$ & $10^8\,\himsun$ & FoF stellar mass threshold to seed BH\\
$\epsilon_r$ & 0.1 & Radiative efficiency of BH \\
$\epsilon_{\rm FB}$ & 0.15 & AGN feedback efficiency \\
$\Delta T_{\rm AGN}$ & $10^{8.5}\,\kelvin$ & AGN feedback temperature\\
\hline
\end{tabular}
\label{tab:parameters}
\end{table*}

\subsection{Star Formation}
\label{sec:starformation}
We assume star formation to occur when the hydrogen number density is higher than the threshold density, $n_{\rm thres}=0.1\,\pcmq$, and the temperature is lower than the threshold temperature, $T_{\rm thres}=10^4\,\kelvin$. 
Each gas particle is allowed to spawn $n_{\rm spawn}$ star particles at most, and the mass of the star particle is $m_* = m_{\rm gas}/n_{\rm spawn}$, where $m_{\rm gas}$ is the mass of the gas particle, and $n_{\rm spawn}=2$.

We use the simple stellar population (SSP) approximation; a star particle represents a cluster of stars with the same metallicity, and their mass function follows an adopted initial mass function (IMF).
For the IMF, its dependence on metallicity and redshift is considered based on the star cluster formation simulation by \citet{2022MNRAS.514.4639C}.
They have investigated IMF at a metallicity range of $10^{-4} \leq Z \leq 10^{-1}$ and redshift range of $0\leq z \leq 20$ and quantified the mass fraction of excess component from a Salpeter-like component,
\begin{equation}
    f_{\rm massive} = 1.07 (1 - 2^x) + 0.04 \times 2.67^x \times z,
\end{equation}
with $x = \min\{1 + \log_{10}(Z/\Zsun), 0\}$, where $z$ is the redshift. The metallicity range investigated by \citet{2022MNRAS.514.4639C} is limited to $Z \leq 0.1\,\Zsun$, and we use the value of $f_{\rm massive}$ at $Z = 0.1\,\Zsun$ for the higher metallicity because IMF at $Z=0.1\,\Zsun$ and $z=0$ is already similar to present day Salpeter-like IMF.
Figure~\ref{fig:imf} shows the variation of IMF depending on $f_{\rm massive}$. We first assume the Chabrier IMF \citep{2003PASP..115..763C} with stellar mass range $0.1\,\Msun < M < 100\,\Msun$, and then add a log-flat component at $5\,\Msun < M < 100\,\Msun$ with mass fraction $f_{\rm massive}$ to allow for a treatment of top-heavy IMF. 

When a star particle is spawned from a gas particle, $f_{\rm massive}$ is computed from the metallicity of the gas particle and the redshift at that time, which is used later for computing energy and metal yield by SNe and AGB stars.
In computing the metal and energy yields, we use the metallicity of the star particle with a floor of $Z_{\rm yield,floor}=0.03\,\Zsun$.
This metallicity floor is introduced to consider the metal enrichment by unresolved early star formation, and the $Z_{\rm yield, floor}$ is similar to the observed metallicity of galaxies with $M_* \sim 10^7\,\Msun$. 

The star formation rate for a gas particle follows the local Schmidt law \citep{1959ApJ...129..243S},
\begin{equation}
    \dot{\rho_*} = \epsilon_* \frac{\rho}{t_{\rm ff}},
    \label{eq:sfr}
\end{equation}
where $t_{\rm ff} = \sqrt{3\pi / 32 G \rho}$ is the local free fall time and $\epsilon_*=0.01$ is the star formation efficiency. 
Star particles are spawned stochastically \citep{1992ApJ...391..502K, 2003MNRAS.339..289S, 2019MNRAS.484.2632S} following the star formation rate.

\subsection{Core Collapse Supernova}
\label{sec:corecollapsesupernova}
When massive stars ($\gtrsim 8\,\Msun$) end their life, their major path is to explode as core-collapse supernovae.
Typically, there is one massive star per $100\,\Msun$ stars, and the kinetic energy of a supernova is $10^{51}\,\erg$; the specific SN energy is $\zeta_{\rm SN}=10^{49}\,\erg\,\Msun^{-1}$.
However, the specific energy can vary due to multiple factors, e.g., stellar IMF, lower and upper limits of SN progenitor mass, and energy per SN.
The specific SN energy adopted in previous works varies by more than a factor of two \citep{2022MNRAS.512..199K}.

Some supernovae, called hypernovae\,(HNe) or superluminous supernovae, have an order of higher explosion energy of $10^{52}\,\erg$.
The number fraction of HNe in a solar metallicity environment is about one percent, but it is observationally suggested that the HN fraction increases at subsolar metallicity \citep{2018SSRv..214...59M, 2019ARA&A..57..305G}.
From a cosmological hydrodynamic simulation of galaxies, \citet{2006ApJ...653.1145K} suggested that the HN fraction of 50\% is necessary to explain the zinc abundance.

\begin{figure}
    \centering
    \begin{minipage}{\columnwidth}
        \includegraphics[width=.8\columnwidth]{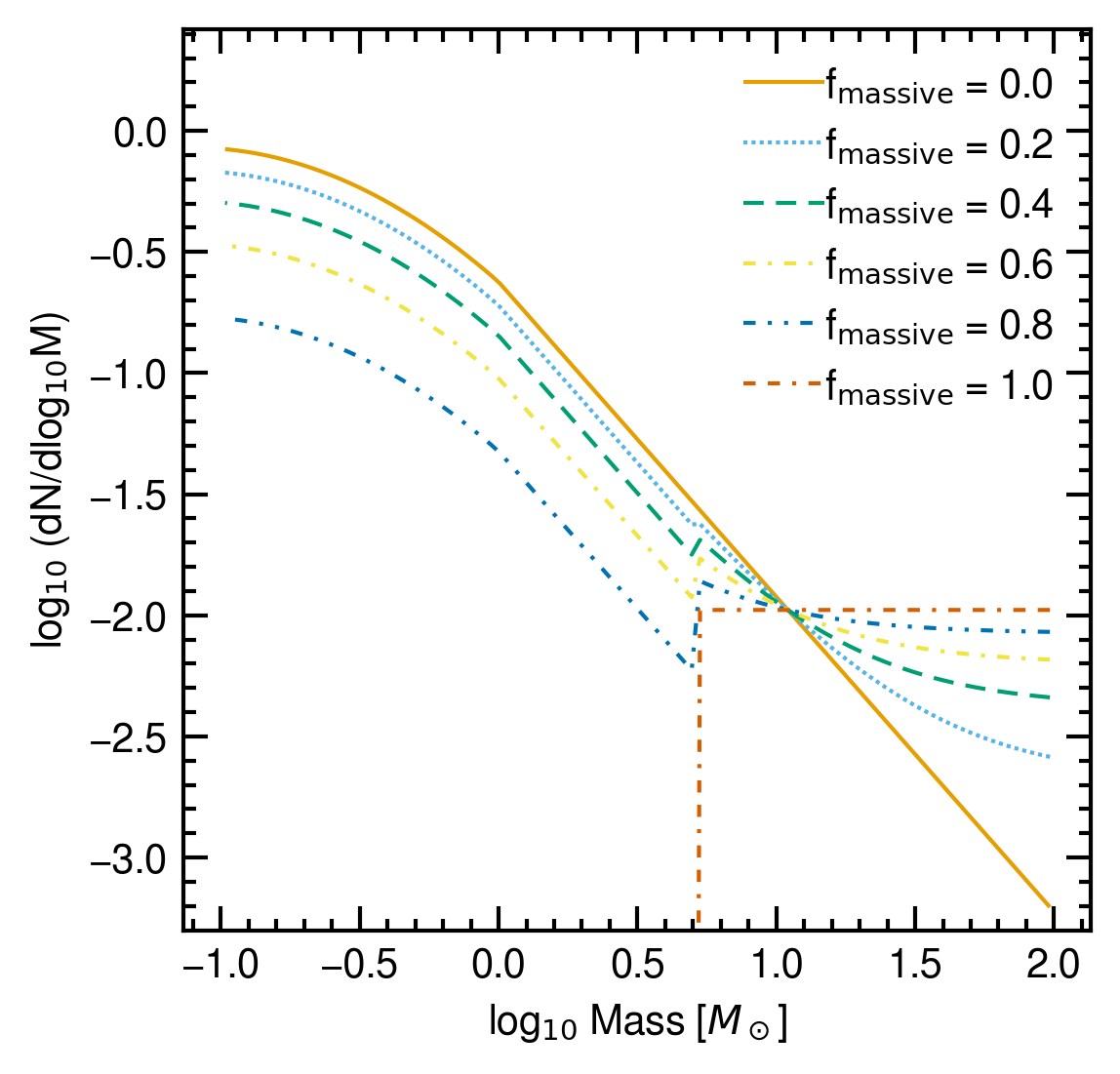}
    \end{minipage}
    
    \begin{minipage}{\columnwidth}
        \includegraphics[width=.8\columnwidth]{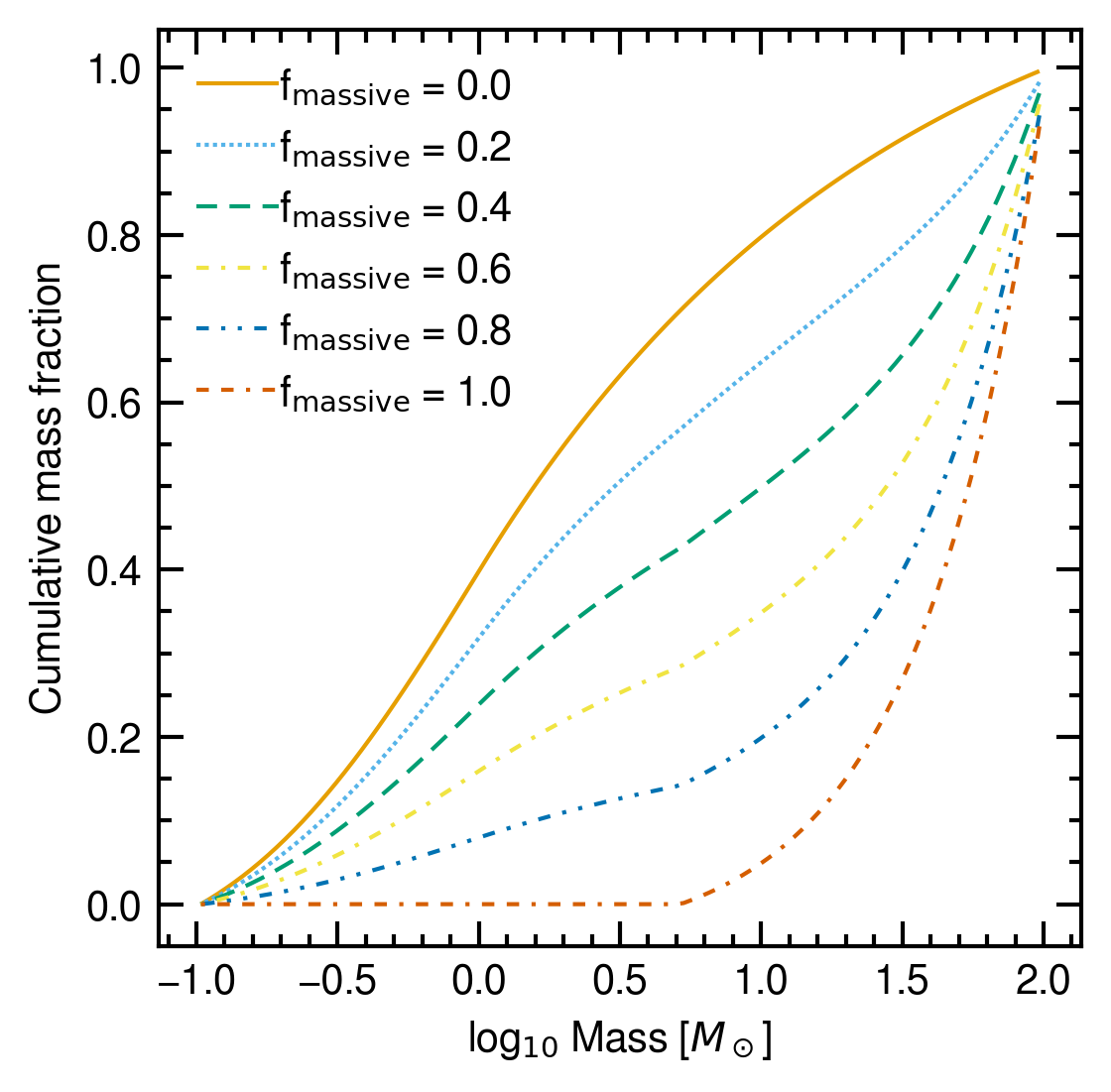}
    \end{minipage}
    
    \begin{minipage}{\columnwidth}
        \includegraphics[width=.8\columnwidth]{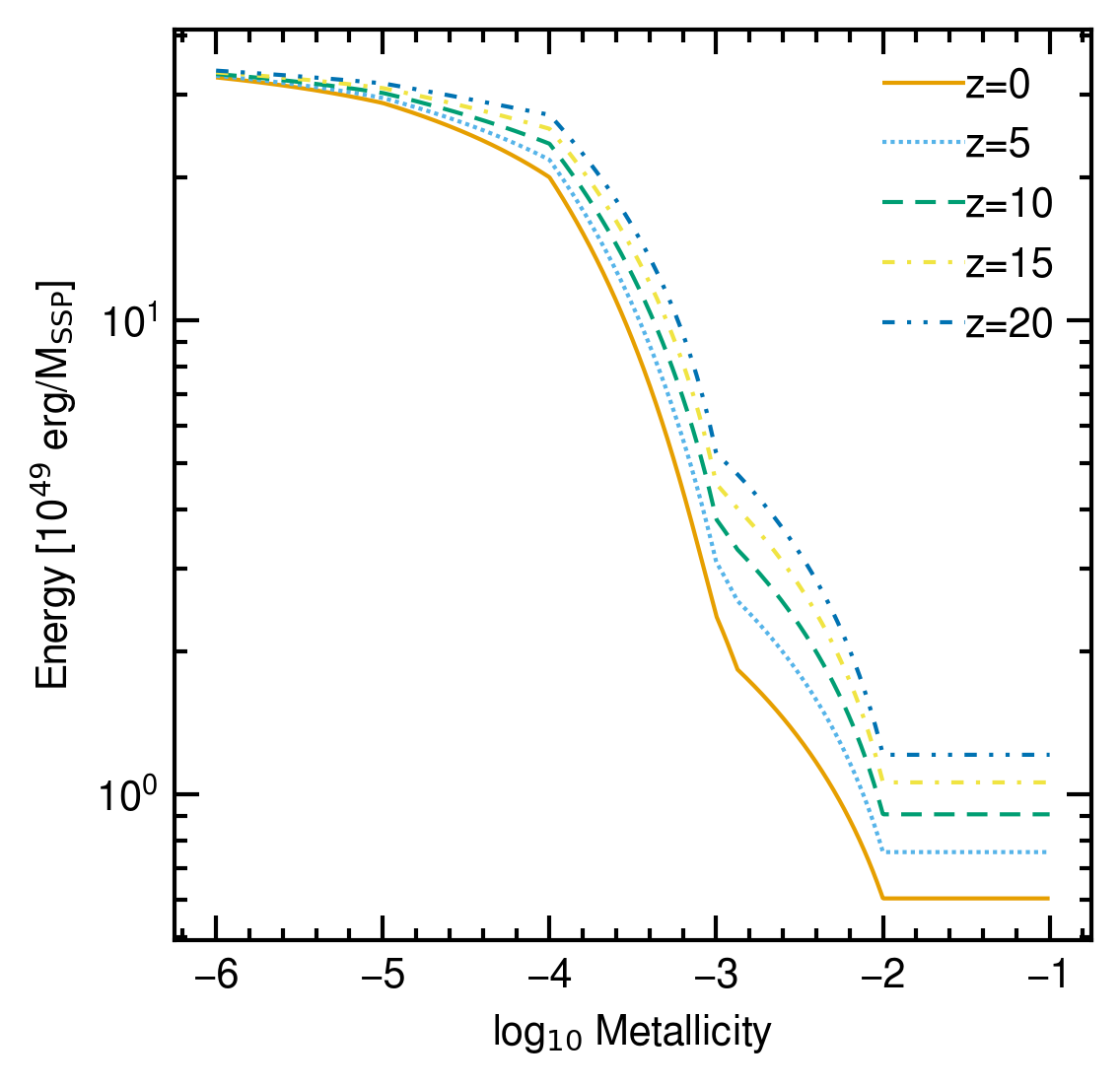}
    \end{minipage}
    
    \caption{{\it Top panel}: Metallicity- and redshift-dependent IMF as a function of stellar mass assumed in this work. {\it Middle}: Cumulative mass fraction of the stars following the IMFs shown in the top panel. {\it Bottom}: Metallicity-redshift dependent specific energy of Type-II supernova.}
    \label{fig:imf}
\end{figure}

In this work, we use the yield model by \citet{2013ARA&A..51..457N}, which covers the progenitor mass range of $13-40\,\Msun$.
We assume that some core-collapse SNe with progenitor masses of $20-40\,\Msun$ explode as HNe, and its number fraction in that mass range is 50\% at $Z<10^{-3}$ and 1\% otherwise.
We generate a yield table as a function of stellar age using the chemical evolution library \textsc{CELib} \citep{2017AJ....153...85S}.
The bottom panel of Figure\,\ref{fig:imf} shows the specific energy of core-collapse SNe as a function of metallicity in our model. 
The specific energy is higher for a low-metallicity star due to the top-heavy IMF and the higher HN fraction.

{The integration of the metallicity- and redshift-dependent top-heavy IMF with the metallicity-dependent HN fraction is the key innovation in {\osakasim}, leading to a significantly higher $\zeta_{\rm SN}$ (by an order of magnitude) at low metallicities and high redshifts compared to previous simulations such as EAGLE and IllustrisTNG. In those simulations, $\zeta_{\rm SN}$ is also introduced as a decreasing function of metallicity to fine-tune the stellar mass function at $z = 0$. However, in their cases, the SN energy is boosted only by a factor of three from the canonical value at low metallicity, based on the idea that the effective SN energy would be higher due to reduced radiative energy loss in a lower metallicity environment \citep{2015MNRAS.450.1937C, 2018MNRAS.473.4077P}. In contrast, our SN feedback model introduces a new consideration, resulting in a significant difference in one of the major feedback parameters. Additionally, we include the metallicity dependence on momentum feedback in our SN feedback model, as detailed later in Section~\ref{sec:mechanicalfeedback}.}

We divide the Type-II SN feedback from a star particle into $n_{\rm event, snii}$ events so that the SN energy and metal yield are deposited gradually rather than instantaneously \citep{2019MNRAS.484.2632S}, and in this work, we adopt $n_{\rm event, snii}=2$.
The SN feedback event occurs following the yield table considering the stellar age.
The energy and metal output of SN is distributed to surrounding $N_{\rm ngb, fb}=8\pm2$ gas particles using the feedback model developed in \citet{2022ApJS..262....9O} with some updates, as we briefly summarize in the following.
The model consists of two components: local mechanical feedback and galactic wind feedback.

\subsubsection{Mechanical Feedback Model}
\label{sec:mechanicalfeedback}
The mechanical feedback model accounts for the momentum injection by SN remnants acquired in the unresolved Sedov--Taylor and pressure-driven snowplow phases.
\citet{2022ApJS..262....9O} have performed three-dimensional hydrodynamic simulations using \textsc{Athena++} code \citep{2020ApJS..249....4S} to investigate the momentum of superbubble formed by clustered SN explosions in a variety of density and metallicity environments with different intervals of SN explosions. 
The terminal momentum obtained in the simulations is normalized over an initial mass function of star clusters and finally translated into the terminal momentum per SN,
\begin{equation}
    \hat{p} = 1.75\times10^5\,\Msun\,\kms\,n_0^{-0.05}\Lambda_{6, -22}^{-0.17},
    \label{eq:momentum_per_SN}
\end{equation}
where $n_0 = \nh/(1\,\pcmq$), and $\Lambda_{6, -22} = \max\{1.9-0.85\,(Z/\Zsun), 1.05\}\times(Z/\Zsun) + 10^{-1.33}$ is the value of cooling function of \citet{1993ApJS...88..253S} at $10^6\,\kelvin$ normalized by $10^{-22}\,\erg\,\psec{\rm cm}^3$.
For every feedback event, the local density and metallicity are computed, and the SN energy is translated to momentum using Eq.\,(\ref{eq:momentum_per_SN}) assuming that the energy per SN is $10^{51}\,\erg$. 

The momentum is distributed to surrounding particles with weights calculated using Voronoi tessellation.
Metals from Type-Ia SN, Type-II SN, and AGB stars are distributed using the same weights.
When coupling the feedback momentum to surrounding particles, we limit the momentum if necessary to ensure manifest energy conservation considering the relative velocity of star and gas particles for individual coupling events \citep[see Appendix B1 of][]{2023MNRAS.519.3154H}.
We did not consider the manifest conservation of total energy and linear momentum of multiple feedback events as discussed in Appendix B3 of \citet{2023MNRAS.519.3154H}, and this would be a task for future development of \textsc{GADGET4-Osaka}, while its effect would be small in our low-resolution simulation where most of SN event is at the momentum-conserving limit on the resolved scale.

\subsubsection{SN-driven Galactic Wind Model}
\label{sec:galacticwindmodel}
The galactic wind model accounts for the hot ($T\gtrsim10^6\,\kelvin$) galactic wind driven by SNe. 
We use the scaling relations by \citet{2020ApJ...903L..34K} to determine mass, energy, and metal loading factors from metallicity and star formation surface density.
The star formation surface density is estimated as $\Sigma_{\rm SFR} = \rho_{\rm SFR} H$, where $\rho_{\rm SFR}$ is the star formation rate density and $H = \rho/|\nabla \rho|$ is the gas scale height obtained using the Sobolev-like approximation.
The model provides the scaling relations for cool ($T\lesssim10^4\,\kelvin$) and hot ($T\gtrsim10^6\,\kelvin$) galactic outflows based on the \textsc{TIGRESS} simulation \citep{2020ApJ...900...61K}, and we employ the model for the hot phase.

In our previous work \citep{2022ApJS..262....9O}, we used a constant entropy wind based on the galactic wind property in the high-resolution dwarf galaxy simulation by \citet{2019MNRAS.483.3363H} assuming a fixed energy loading factor $\eta_e = 0.7$.
By updating our model as described above, we no longer have to assume $\eta_e$, thus reducing one free parameter. 

Our simulation has the mechanical feedback model, which drives the cool wind, and we did not use the wind model for the cool phase. The galactic wind model for the hot phase is necessary for low-resolution simulations that cannot resolve the Sedov--Taylor phase because the mechanical feedback model is for the snowplow phase where the SN bubble has cooled down, and treatment to avoid overcooling is necessary to produce hot galactic wind \citep{2022ApJS..262....9O}. 

We follow the \textsc{Twind} sampling procedure described in Appendix B of \citet{2020ApJ...903L..34K} to sample wind particles and determine their temperature, metallicity, and wind velocity except for the first step. 
For the first step, we obtain the mass of the hot wind as $M_{\rm hot} = \tilde{\eta}^{\rm hot}_{M} (m_*/n_{\rm event, snii})(\zeta_{\rm SN}/\zeta_{\rm TIGRESS})$, where $\tilde{\eta}^{\rm hot}_{M}$ is the mass loading factor obtained as a function of $\Sigma_{\rm SFR}$.
The second term represents the stellar mass that contributes to one feedback event.
The third term is a factor considering the difference in the energy yield, where $\zeta_{\rm SN}$ and $\zeta_{\rm TIGRESS} $ are the specific SN energy adopted in our simulation and the TIGRESS framework \citep{2017ApJ...846..133K, 2020ApJ...903L..34K}. 
The specific SN energy adopted in our simulation is redshift- and metallicity-dependent as shown in Fig.~\ref{fig:imf}, and $\zeta_{\rm TIGRESS} = 1.05\times10^{49}\,\erg\,\Msun^{-1}$.

The sampled particles are flagged as wind particles and kicked at the wind velocity in random directions isotropically.
We assume that wind particles represent the hot galactic wind flowing through unresolved low-density channels, and we disable their hydrodynamical interaction and cooling to avoid the wind particles being affected by ISM. 
We recouple the wind particle to hydrodynamics and enable cooling after its density falls below 0.05 times the threshold density of star formation or 0.025 times the current Hubble time has elapsed, following the criteria used in the IllustrisTNG model \citep{2018MNRAS.473.4077P}.

\subsection{Type-Ia Supernova}
For Type-Ia SNe, the element yield table by \citet{2013MNRAS.429.1156S} is used in combination with a power-law type delay-time distribution function of $\Psi(t) = 4\times10^{-13}\,{\rm SN}\,{\rm yr}^{-1}\,\Msun^{-1}\,(t / 1\,{\rm Gyr})^{-1}$ with a minimum delay of 40 Myr \citep{2012PASA...29..447M}.
We divide the Type-Ia SN feedback from a star particle into $n_{\rm event, snia}=8$ events.
The energy per Type-Ia SN is fixed to $10^{51}\,\erg$, and we use Eq.\,(\ref{eq:momentum_per_SN}) to convert the energy to feedback momentum. We did not consider thermal galactic wind feedback for Type-Ia SNe because their degree of clustering is low, and they are not considered to be the major driver of the galactic wind.

\subsection{AGB Stars}
For AGB stars, two yield tables are combined to cover the mass range of 1--8 $\Msun$; \citet{2010MNRAS.403.1413K} table is used in $M_*/M_\odot \in [1, 6]$, \citet{2014MNRAS.437..195D} table in $M_*/M_\odot \in [7, 8]$, and we linearly interpolate between them.
We do not consider the mechanical feedback due to the stellar wind of AGB stars.
We divide the AGB feedback from a star particle into $n_{\rm event, agb}=8$ events.

\subsection{Black Hole Physics}
For the supermassive black holes (BHs), we largely follow the model used in the EAGLE project \citep{2015MNRAS.446..521S, 2015MNRAS.450.1937C}. The model consists of seeding using the Friend-of-Friends (FoF) halo finder, torque-limited Bondi-Hoyle accretion, and thermal quasar feedback, as summarized below.
\subsubsection{Seeding \& Repositioning}
We regularly run the FoF halo finder with an interval of $\Delta \ln a = 0.005$, where $a = 1/(1+z)$ is the scale factor, and convert the gas particle with a minimum gravitational potential in the halo to a BH particle if the total mass of the halo is larger than $M_{\rm seeding, FoF}=10^{10}\,\himsun$, stellar mass in the halo is larger than $M_{\rm seeding, star}=10^8\,\himsun$, and does not contain BHs.

If a FoF halo contains BH particles and the BHs are within three times the gravitational softening length of BH particles from the star particle with minimum potential, the BH particles are repositioned and assigned the velocity of the FoF halo.

\subsubsection{Accretion}
The mass accretion rate onto BH is computed by the Eddington-limited Bondi rate. The Bondi rate \citep{1952MNRAS.112..195B} is
\begin{equation}
    \dot{M}_{\rm Bondi} = \frac{4 \pi G^2 M_{\rm BH}^2  \rho}{(c_s^2 + v_{\rm BH}^2)^{3/2}},
\end{equation}
where $G$ is the gravity constant, $M_{\rm BH}$ is the BH mass, and $v_{\rm BH}$ is the velocity of the BH relative to the surrounding medium. We set $v_{\rm BH}$ to zero because the velocities of BH particles are set by hand when repositioned. 
In computing the sound speed $c_s$, we assume the effective equation of state (EoS) used in the EAGLE project, i.e., $c_s = \sqrt{\gamma (\kB/\mu_{\rm a} m_{\rm p}) T_{\rm eos}}$, where $\gamma=5/3$ is the adiabatic index, $\kB$ is the Boltzmann constant, $\mu_{\rm a} = 1.2285$ is the mean molecular weight of primordial atomic gas, $m_{\rm p}$ is the proton mass, and $T_{\rm eos} = 8\times10^3\,\kelvin\,(\nh / 0.1\,\pcmq)^{1/3}$ is the temperature derived from the EoS. 
The EoS is used only to compute the BH accretion rate for consistency with the accretion model of the EAGLE simulation.

The Eddington accretion rate is
\begin{equation}
    \dot{M}_{\rm Eddington} = \frac{4 \pi G M_{\rm BH} m_{\rm p}}{\epsilon_r c \sigma_{\rm T}},
\end{equation}
where $\epsilon_r = 0.1$ is the radiative efficiency of BH, $c$ is the speed of light, and $\sigma_{\rm T}$ is the Thomson cross section. 
The accretion rate is the minimum of the Bondi rate and the Eddington rate
\begin{equation}
    \dot{M}_{\rm acc} = \min(\dot{M}_{\rm Eddington}, C \dot{M}_{\rm Bondi}),
    \label{eq:cvisc}
\end{equation}
where $C = \min(C_{\rm visc}^{-1} (c_s/v_\phi)^3, 1)$ is a factor to limit the Bondi rate considering the angular momentum and the viscosity of the accretion disk on a subgrid scale \citep{2015MNRAS.454.1038R}. 
We set the model parameter $C_{\rm visc}$ to $200 \pi$ in our fiducial run, and $v_\phi$ is the rotation speed of the gas around the BH.

We use the subgrid BH mass to follow the growth of BH \citep{2005MNRAS.361..776S}.
Each BH particle has particle mass, $m_{\rm part}$, and subgrid BH mass, $M_{\rm BH}$; the former is used in the computation of gravitational force and potential, and the latter is used for the calculation of the accretion rate onto the black hole.
The mass growth rate of the BH is 
\begin{equation}
    \dot{M}_{\rm BH} = (1 - \epsilon_r) \dot{M}_{\rm acc}.
\end{equation}
The BH particles swallow the surrounding gas particles stochastically following their subgrid BH mass.
For each gas particle $j$ around the BH, we compute a probability
\begin{equation}
    p_{j, {\rm acc}} = \frac{w_{j} \Delta m }{\rho},
\end{equation}
where $w_j$ is the kernel weight of the gas particle relative to the BH, $\Delta m = \max(M_{\rm BH} - m_{\rm part}, 0)$ is the mass increment from the BH particle mass to the subgrid mass, and $\rho$ is the gas density at the position of the BH. We then draw a uniform random number $x_j \in [0, 1]$, and the BH absorbs the gas particle with conserving mass and momentum if $x_j < p_{j, {\rm acc}}$.

\subsubsection{Feedback}
\label{sec:AGNfeedback}
The AGN feedback is modeled by stochastic thermal energy injection. 
Each BH particle has an energy reservoir, and we increase it at a rate of $\dot{E}_{\rm res} = \epsilon_r \epsilon_{\rm FB} \dot{M}_{\rm acc} c^2$, where $\epsilon_{\rm FB} = 0.15$ is the efficiency of converting AGN radiation to thermal energy by radiative pressure.
When $E_{\rm res}$ is large enough to increase the temperature of a gas particle by $\Delta T_{\rm AGN}$, the BH is allowed to distribute the feedback energy stochastically.
Similarly to the treatment of accretion, for each gas particle $j$ around the BH, we compute a probability
\begin{equation}
    p_{j, {\rm FB}} = \frac{(\gamma - 1) \mu_{\rm i} m_{\rm p} w_j E_{\rm res} }{\kB \Delta T_{\rm AGN} \rho},
\end{equation}
where $\mu_{\rm i} = 0.588$ is the mean molecular weight of a primordial ionized gas,
and draw a uniform random number $y_j \in [0, 1]$.
The thermal energy of 
\begin{equation}
    E_{j, {\rm FB}} = \frac{m_j \kB \Delta T_{\rm AGN}}{(\gamma - 1) \mu m_p}
\end{equation}
is added to the gas particle if $y_j < p_{j, {\rm FB}}$, and then we reduce $E_{\rm res}$ by $E_{j, {\rm FB}}$. 

\subsubsection{Timestepping}
The timestep size of a black hole is limited to constrain the accreting mass and feedback energy per time step.
The accretion timestep is determined to limit the mass growth rate of a black hole per time step to less than 10\% and the number of SPH particles swallowed by a black hole to one:
\begin{equation}
    \Delta t_{\rm acc} = \min\left\{0.1 \frac{M_{\rm BH}}{\dot{M}_{\rm BH}}, \frac{\bar{m}_{\rm gas}}{\dot{M}_{\rm BH}}\right\},
\end{equation}
where $\bar{m}_{\rm gas}$ is the mean gas particle mass.
The feedback timestep is determined as
\begin{equation}
    \Delta t_{\rm FB} = 0.3 \frac{N_{\rm ngb, BH} \bar{m}_{\rm gas} \kB \Delta T_{\rm AGN}}{(\gamma - 1) \mu m_p \dot{E}_{\rm res}},
\end{equation}
such that the fraction of neighboring SPH particles that receive feedback energy is less than 30\% within the kernel. 
We use the shortest of $\Delta t_{\rm acc}$, $\Delta t_{\rm FB}$, and the gravitational timestep size of the black hole particle.

\subsection{Runs}
\label{sec:runs}

\begin{table*}
    \caption{List of simulations. The columns list (1) the run name; (2) the box size in comoving Mpc, $L_{\rm box}$; (3) the number of particles employed in the run, $N_{\rm particle}$; (4) the subgrid viscous parameter for BH accretion, $C_{\rm visc}$; (5) the subgrid mass of BH seed, $m_{\rm seed}$; (6-8) the checklist of feedback modules activated in the run, (6) the mechanical feedback from core-collapse SNe (see Section~\ref{sec:mechanicalfeedback}), (7) the SN-driven galactic wind model (see Section~\ref{sec:galacticwindmodel}), and (8) AGN feedback (see Section~\ref{sec:AGNfeedback}); (9) the stellar IMF; (10) the number fraction of HNe.}
    \centering
    \begin{tabular}{lccccccccc}
    \hline
        &  &  &  &  & \multicolumn{3}{c}{Feedback Type} & & \\ \cmidrule{6-8}
       Name & $L_{\rm box}$ & $N_{\rm particle}$ & $C_{\rm visc}$ & $m_{\rm seed}$ & SN Mechanical & SN GalWind & AGN & Stellar IMF & HN fraction \\
       (1) & (2) & (3) & (4) & (5) & (6) & (7) & (8) & (9) & (10)\\
       \hline
       Fiducial  & $50$ & $2\times512^3$ & $200\pi$ & $1\times10^5$ & \checkmark & \checkmark & \checkmark & variable\footnote{The metallicity- and redshift-dependent IMF (see Section~\ref{sec:starformation})} & $Z$-dependent\footnote{The redshift-dependent HN fraction (see Section~\ref{sec:corecollapsesupernova})}\\
       NoSNGalWind  & $50$ & $2\times512^3$ & $200\pi$ & $1\times10^5$ &  \checkmark &  & \checkmark & variable & $Z$-dependent\\
       AGNonly & $50$ & $2\times512^3$  & $200\pi$ & $1\times10^5$ &  & & \checkmark & variable & $Z$-dependent\\
       SNonly     & $50$ & $2\times512^3$ & $200\pi$ & $1\times10^5$ & \checkmark & \checkmark & & variable & $Z$-dependent \\
       {NoZdepSN} & $50$ & $2\times512^3$ & $200\pi$ & $1\times10^5$ &  \checkmark & \checkmark & \checkmark & Chabrier & 0.01 \\
       NoFB      & $50$ & $2\times512^3$ & $200\pi$ & $1\times10^5$ & & &  & variable & $Z$-dependent\\
       LowCvisc & $50$ & $2\times512^3$ & $2\pi$ & $1\times10^5$ &  \checkmark & \checkmark & \checkmark & variable & $Z$-dependent\\
       LowCviscLowMseed & $50$ & $2\times512^3$ & $2\pi$ & $1\times10^4$ &  \checkmark & \checkmark & \checkmark & variable & $Z$-dependent\\
       L25N512  & $25$ & $2\times512^3$ & $200\pi$ & $1\times10^5$ &  \checkmark & \checkmark & \checkmark & variable & $Z$-dependent\\
       L25N256    & $25$ & $2\times256^3$ & $200\pi$ & $1\times10^5$ &  \checkmark & \checkmark & \checkmark & variable & $Z$-dependent\\
       L25N128  & $25$ & $2\times128^3$ & $200\pi$ & $1\times10^5$ &  \checkmark & \checkmark & \checkmark & variable & $Z$-dependent\\
       \hline
    \end{tabular}
    \label{tab:sims}
\end{table*}

We perform simulations with varied feedback settings, black hole accretion parameters, and resolution as summarized in Table \ref{tab:sims}.

The Fiducial run considers the two-component SN feedback and AGN feedback.
The NoSNGalWind run does not use the SN-driven hot galactic wind model. It is presented to show the effect of explicit modeling of the galactic wind in comparison to the Fiducial run.
The AGNonly and SNonly runs only consider AGN and SN feedback, respectively, and are performed to investigate the impact of these feedbacks.
{In the NoZdepSN run, the stellar IMF is fixed to the Chabrier IMF, and the hypernova fraction is set to 0.01, independent of metallicity and redshift.
This is different from other runs, where the specific energy of Type-II SN is higher for lower metallicity stars as shown in Fig.~\ref{fig:imf} due to metallicity- \& redshift-dependent IMF and hypernova fraction.}
The NoFB run does not consider any feedback and is presented as a baseline to show the impact of feedback in comparison to other runs.

The LowCvisc and LowCviscLowMseed runs adopt black hole accretion parameters different from the Fiducial run.
In these two runs, we adopt $C_{\rm visc}=2\pi$ (same value as the fiducial EAGLE simulation), which is 100 times smaller than the one adopted in the Fiducial run.
The parameter $C_{\rm visc}$ modulates the limiting value of parameter $C$ in Eq.~\ref{eq:cvisc} through the accretion suppression factor of $C_{\rm visc}^{-1} (c_s/v_\phi)^3$, which is usually less than unity.
The lower $C_{\rm visc}$ results in a higher suppression factor \citep{2015MNRAS.454.1038R, 2015MNRAS.446..521S, 2015MNRAS.450.1937C},
which allows BHs to accrete at near Bondi rate, hence faster and earlier growth at $M_* \simeq 10^9-10^{10}\,\Msun$ as we will see in Section~\ref{sec:result_bh}.
The LowCviscLowMseed adopts the BH seed mass of $M_{\rm seed}=10^4\,\himsun$, 10 times smaller than that used in the other two runs.

The L25N512, L25N256, and L25N128 runs have different mass resolutions and are performed to investigate the dependence on resolution.
The size of the simulation box of the L25N256 run is $25\,\hicmpc$, and the number of employed particles is $2\times256^3$, as the name indicates. The L25N256 run has the same resolution as the Fiducial run (i.e., L50N512). 
The L25N512 and L25N128 have eight times better and worse mass resolution than the L25N256 run.
The same configurations and parameters are employed in the three runs except for the gravitational softening length.
In the L25N256 run, we use the same gravitational softening length as the Fiducial run, and we set it to twice as small and as large in the L25N512 and L25N128 runs.

\subsection{Data Analysis}
\label{sec:method_analysis}
Simulation data is analyzed by on-the-fly friends-of-friends (FoF) and the \textsc{subfind} group and substructure finder \citep{2001MNRAS.328..726S, 2021MNRAS.506.2871S}.
For visualization and further data analysis, we generate a uniform three-dimensional cartesian mesh covering the entire simulation box with $1024^3$ voxels, assign particle data, and take projections on the fly. The cloud-in-cell (CIC) assignment is used for dark matter and star particles, and the SPH kernel weight is used for gas particles.
The SPH particle data is assigned to the mesh, conserving the total mass, metal mass, and internal energy.
For only data analysis purposes, we set a floor in the SPH smoothing length at $(\sqrt{3}/2) L_{\rm voxel}$, where $L_{\rm voxel}$ is the size of a voxel, so that there are some voxel centers within the smoothing length from each SPH particle.
We then assign the mass of $m_i W_{ij}(h_i)/\sum_k^N W_{ik}(h_i)$ to the $j$ -th voxel of the $i$-th particle, where $m_i$ and $h_i$ are the mass and smoothing length of the $i$-th particle, and $W_{ij}(h_i)=W(r_{ij}; h_i)$ is the kernel weight, $r_{ij}$ being the distance from the $i$-th particle to the center of $j$-th voxel.
The metal mass and internal energy are assigned to voxels in the same manner, and the metallicity and temperature of each voxel are calculated using the mass, metal mass, and internal energy of the voxel.
This mesh generator module is useful for the post-process analysis of the IGM presented in Section~\ref{sec:IGM}.

\begin{figure*}
    \centering  
    \begin{interactive}{animation}{fig2_anim}
    \includegraphics[width=.9\textwidth]{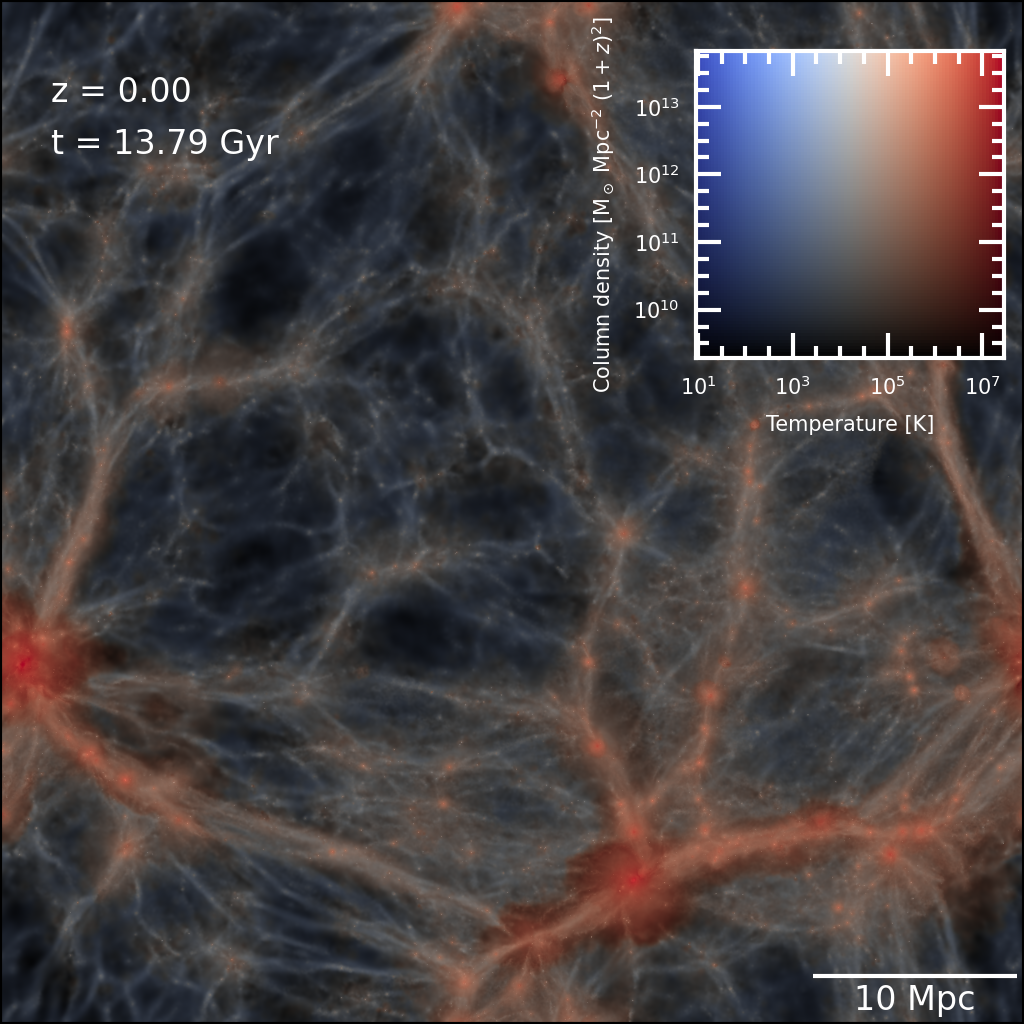}
    \end{interactive}
    \caption{Projected image through a $6.25\,\hicmpc$ slice of the Fiducial run. The color and intensity indicate temperature and density, as shown by the color map at the top right corner of the image. An animated version of this figure, showing the evolution from $z$ = 39 to 0, is available in the HTML version of this article.}
    \label{fig:L50fiducial_projection}
\end{figure*}

Figure~\ref{fig:L50fiducial_projection} shows the density projection image of the Fiducial run at $z=0$. The baryonic cosmic web, composed of intermediate-temperature ($T\sim10^4\,\kelvin$) filaments and high-temperature nodes, is apparent.  The rest of the cosmic volume is filled with low-temperature voids.
\footnote{The movie is available at \url{https://www.yurioku.com/research/}}

\section{Galaxy Statistics}
\label{sec:galaxystatistics}
This section compares the statistical properties of galaxies formed in our simulations. In the following subsections, we compare simulations focusing on the feedback model variation, BH model parameters, resolution, and SN feedback energy variation.
We examine the stellar mass, star formation rate (SFR), gas-phase metallicity, and the most massive black hole mass in galaxies, identified as subhalos by the \textsc{subfind} algorithm.
The stellar mass presented in the following is the mass of stars within twice half-stellar-mass radius; for each collection of stellar particles in a subhalo, we compute the radius of a sphere enclosing half of its mass and then obtain the stellar mass within the twice half-stellar-mass radius.
The SFR of a galaxy is the sum of the gas particles' SFR, computed using Eq.\,\ref{eq:sfr}, in the subhalo.
The gas-phase metallicity of a galaxy is an SFR-weighted average of the metallicity of gas particles in the subhalo, intended for comparison with observations where metallicity is measured using nebular lines from ionized regions around young stars in star-forming regions.
In the following figures, we show various quantities as a function of galaxy stellar mass above $M_*=10^{8.5}\,\Msun$ because the mass of each stellar particle is $9.1\times10^6\,\Msun$.

\subsection{Feedback Model Variation}

\begin{figure*}
    \centering
    \includegraphics[width=\textwidth]{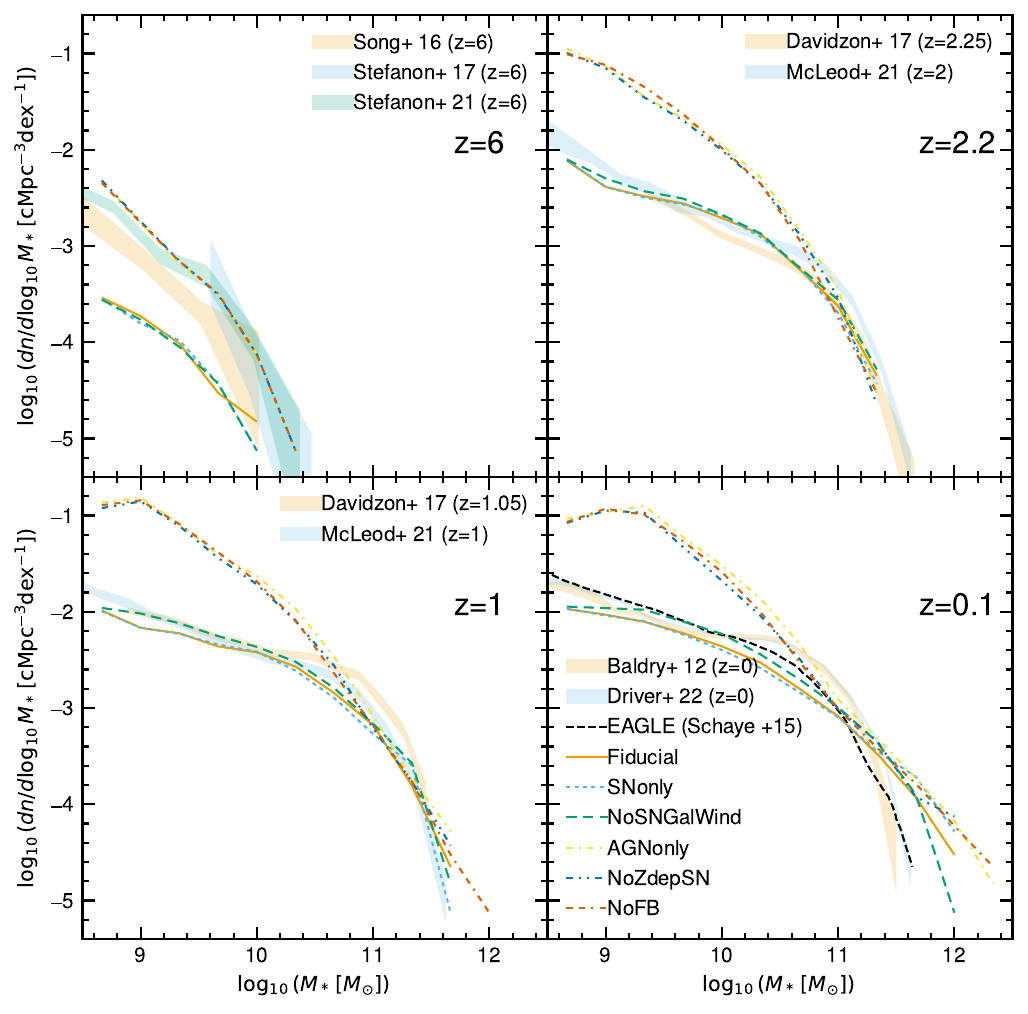}
    \caption{Galaxy stellar mass function at $z=0.1$, 1, 2.2, 6. The black dashed line in the bottom right panel is for the Ref run of EAGLE simulation \citep{2015MNRAS.446..521S}. 
    The shaded regions are the observational constraints on SMF from \citet{2016ApJ...825....5S, 2017ApJ...843...36S, 2021ApJ...922...29S} at $z=6$, \citet{2021MNRAS.503.4413M} at $z=1$ and 2, \citet{2017A&A...605A..70D} at $z=2.25$, and \citet{2012MNRAS.421..621B} and \citet{2022MNRAS.513..439D} around $z=0.1$.}
    \label{fig:SMF}
\end{figure*}

Figure~\ref{fig:SMF} shows the galaxy stellar mass function\,(SMF) at $z=6$, 2.2, 1, and 0.1. 
The shaded regions are the observational constraints on SMF from \citet{2016ApJ...825....5S, 2017ApJ...843...36S, 2021ApJ...922...29S} at $z=6$, \citet{2021MNRAS.503.4413M} at $z=1$ and 2, \citet{2017A&A...605A..70D} at $z=2.25$, and \citet{2012MNRAS.421..621B} and \citet{2022MNRAS.513..439D} around $z=0.1$.

The Fiducial, SNonly, and NoGalWind runs show suppressed star formation compared to the runs without SN mechanical feedback (AGNonly and NoFB) {or the run with weaker SN feedback (NoZdepSN)} at $M_*<10^{11}\,\Msun$, leading to a better consistency with the observation at $z=2.2$ and 1.
This indicates that the SN mechanical feedback {boosted by top-heavy IMF and hypernova} is the dominant feedback source for the low-mass galaxies.
We discuss the implication of the discrepancy at $z=6$ in Section~\ref{sec:summary}.

At $z=0.1$, the Fiducial run nicely reproduces the low-mass end of the observed SMF.
Both our simulations and the EAGLE simulation decrease towards a higher mass end with a slightly steeper slope than the observed exponential cutoff at $M_*\sim 10^{11.5}\,\Msun$, but capture the cutoff mass range relatively well.

{In the Fiducial run, we used the metallicity- and redshift-dependent SN energy yield model, which has increasing specific SN energy towards lower SN progenitor metallicity as shown in Fig.~\ref{fig:imf}.
In the NoZdepSN run, the specific SN energy is fixed to $\zeta_{\rm SN}\simeq6\times 10^{48}\,\erg\,\Msun^{-1}$.
One can see that the NoZdepSN run completely overpredicts the number of low-mass galaxies at $z<2.2$.
This indicates that the SN feedback with the above constant $\zeta_{\rm SN}$ is not strong enough to suppress star formation and reproduce the observed SMF.}

\begin{figure*}
    \centering
    \includegraphics[width=\textwidth]{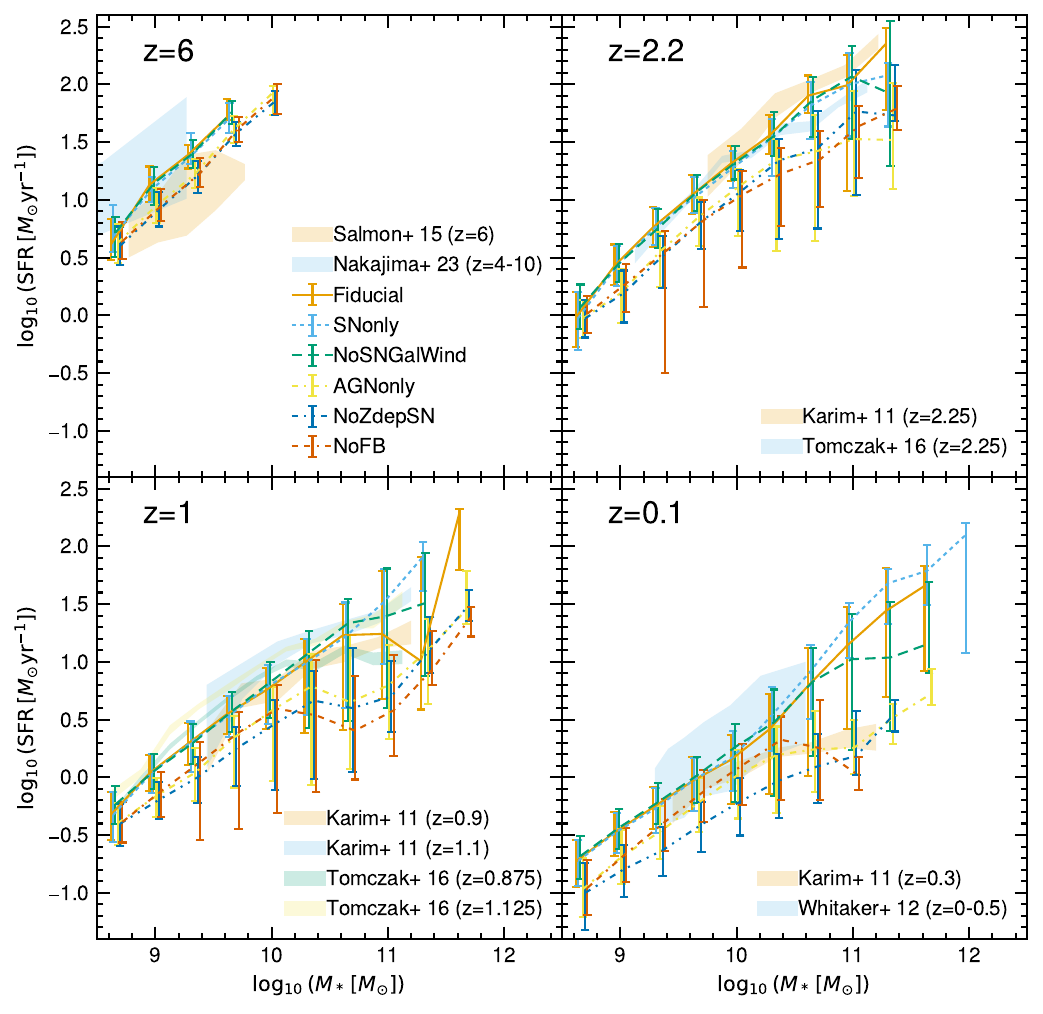}
    \caption{Redshift evolution of the star formation main sequence of the simulated galaxies with sSFR $> 10^{-2}\,{\rm Gyr}^{-1}$. The lines and error bars show the median SFR values and 16th and 84th percentile for each stellar mass bin in our simulations.  The shaded regions are  the observational constraints from \citet{2015ApJ...799..183S} at $z=6$, \citet{2023ApJS..269...33N} at $z=$4-10, \citet{2011ApJ...730...61K} at $z=2.25$, 1.1, 0.9, and 0.3, \citet{2016ApJ...817..118T} at $z=2.25$, 1.125, and 0.875, and \citet{2012ApJ...754L..29W} at $z=$0-0.5.}
    \label{fig:SFMS}
\end{figure*}

Figure~\ref{fig:SFMS} shows the star-formation main sequence (SFMS) at $z=6$, 2.2, 1, and 0.1.
The lines and error bars show the median SFR values and 16th and 84th percentile for each stellar mass bin in our simulations.  The shaded regions are  the observational constraints from \citet{2015ApJ...799..183S} at $z=6$, \citet{2023ApJS..269...33N} at $z=$4-10, \citet{2011ApJ...730...61K} at $z=2.25$, 1.1, 0.9, and 0.3, \citet{2016ApJ...817..118T} at $z=2.25$, 1.125, and 0.875, and \citet{2012ApJ...754L..29W} at $z=0-0.5$.
We display only the star-forming galaxies, selected by a criterion sSFR $> 10^{-2}\,{\rm Gyr}^{-1}$, because the observed data are also for star-forming galaxies. 

At $z=6$, the simulations show similar SFMS with each other.
The runs without stellar feedback {or with weaker SN feedback} show lower SFR than the observed SFMS for a given stellar mass at lower redshifts.
This indicates that stellar feedback is necessary to sustain star formation activity by driving interstellar turbulence and galactic wind to delay the depletion of gas in the dark matter halo and produce star-forming galaxies at $z=0$.
Runs with stellar feedback show a nice agreement with observations.

\begin{figure*}
    \centering
    \includegraphics[width=\textwidth]{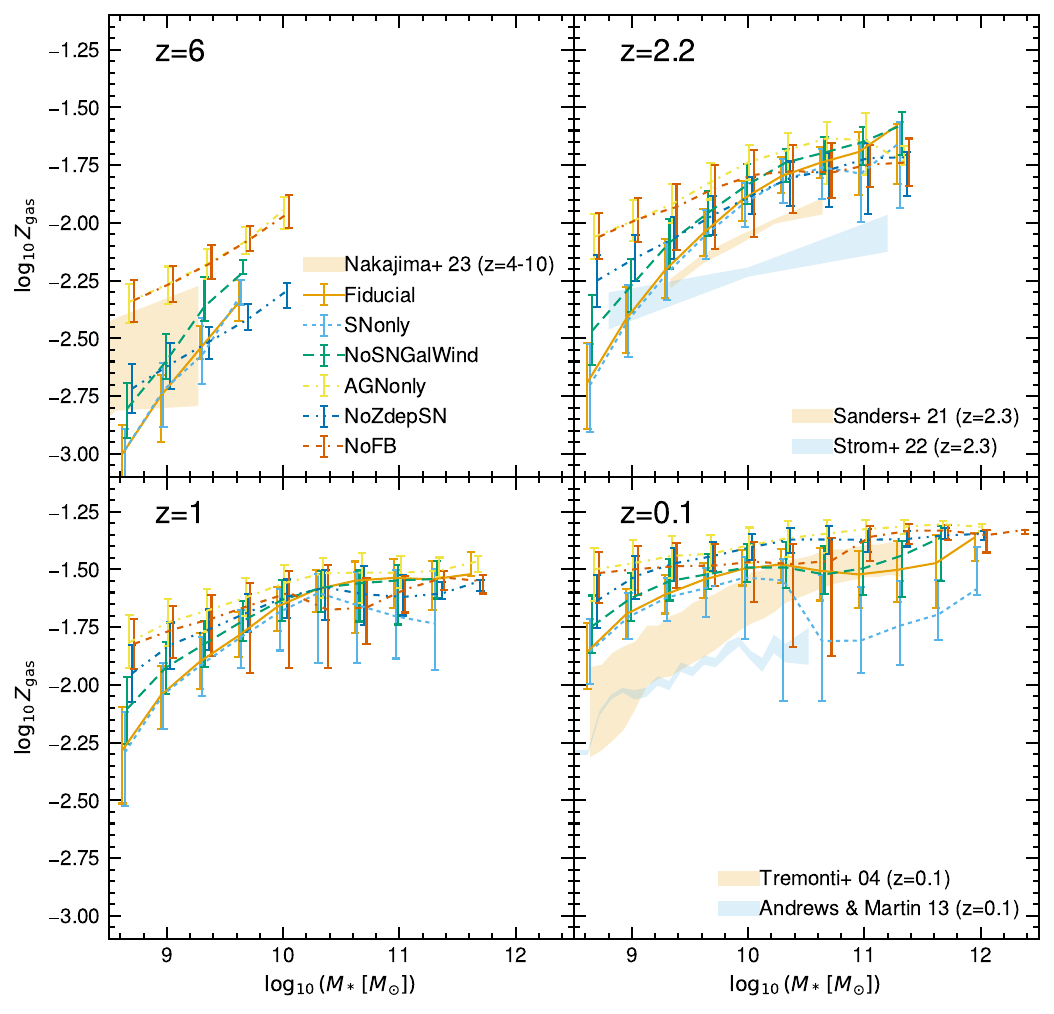}
    \caption{Redshift evolution of the mass--metallicity relation.  The lines and error bars show the median metallicity values and 16th and 84th percentile for each stellar mass bin in our simulations.
    The shaded regions are observational constraints from \citet{2023ApJS..269...33N} at $z=$4-10, \citet{2021ApJ...914...19S} and \citet{2022ApJ...925..116S} at $z=2.3$, and \citet{2013ApJ...765..140A} and \citet{2004ApJ...613..898T} at $z=0.1$.}
    \label{fig:MZR}
\end{figure*}

Figure~\ref{fig:MZR} shows the mass--metallicity relations (MZR) at $z=6$, 2.2, 1, and 0.1.
The metallicity of simulated galaxies is computed with the total metal mass in the gaseous phase.
The lines and error bars show the median metallicity values and 16th and 84th percentile for each stellar mass bin in our simulations.
The shaded regions are observational constraints from \citet{2023ApJS..269...33N} at $z=$4-10, \citet{2021ApJ...914...19S} and \citet{2022ApJ...925..116S} at $z=2.3$, and \citet{2013ApJ...765..140A} and \citet{2004ApJ...613..898T} at $z=0.1$.
The galaxies in the runs without stellar feedback have higher metallicity at $M_* \lesssim 10^9\,\Msun$, which indicates that stellar feedback expels metals from galaxies.

One can see that the NoSNGalWind run has slightly higher metallicity than the Fiducial run. 
This is because of the metal reduction by the galactic wind in the Fiducial run.
The typical metal loading factor of our galactic wind model is 0.3, corresponding to 0.15 dex, which can explain the difference between the NoSNGalWind and Fiducial run.
The difference among the models is small at $M_* \gtrsim 10^{10}\,\Msun$ because the feedback effect is weaker at the massive end, where the gravitational binding energy is stronger.

The metallicity of simulated galaxies of $M_* \sim 10^{10}\,\Msun$ is 0.2 dex higher than the MZR by \citet{2021ApJ...914...19S} at $z=2.2$, and 0.4 dex higher than the MZR by \citet{2013ApJ...765..140A} at $z=0.1$.
{This can be due to either inefficient metal ejection by the galactic wind or insufficient gas mass fraction in the simulated galaxies at low redshift. 
A low gas fraction can be caused by strong feedback suppressing the accretion of low-metallicity gas or active star formation converting gas into stars.
The metallicity increases as Type-Ia SNe and AGB stars slowly supply metals into ISM with certain time delays ($\sim$100 Myr) without being affected by instantaneous feedback by Type-II SNe. 
We will investigate the galactic outflow properties in our simulations by comparing them with recent IFU data at $z=0$-2, as well as the gas fraction and the metal abundance pattern in ISM in our future work.}

{The NoZdepSN run has lower metallicity than the NoFB run, while they agree in SMF and SFMS. This is because the top-heavy IMF is not adopted in the NoZdepSN run.}

\begin{figure*}
    \centering
    \includegraphics[width=\textwidth]{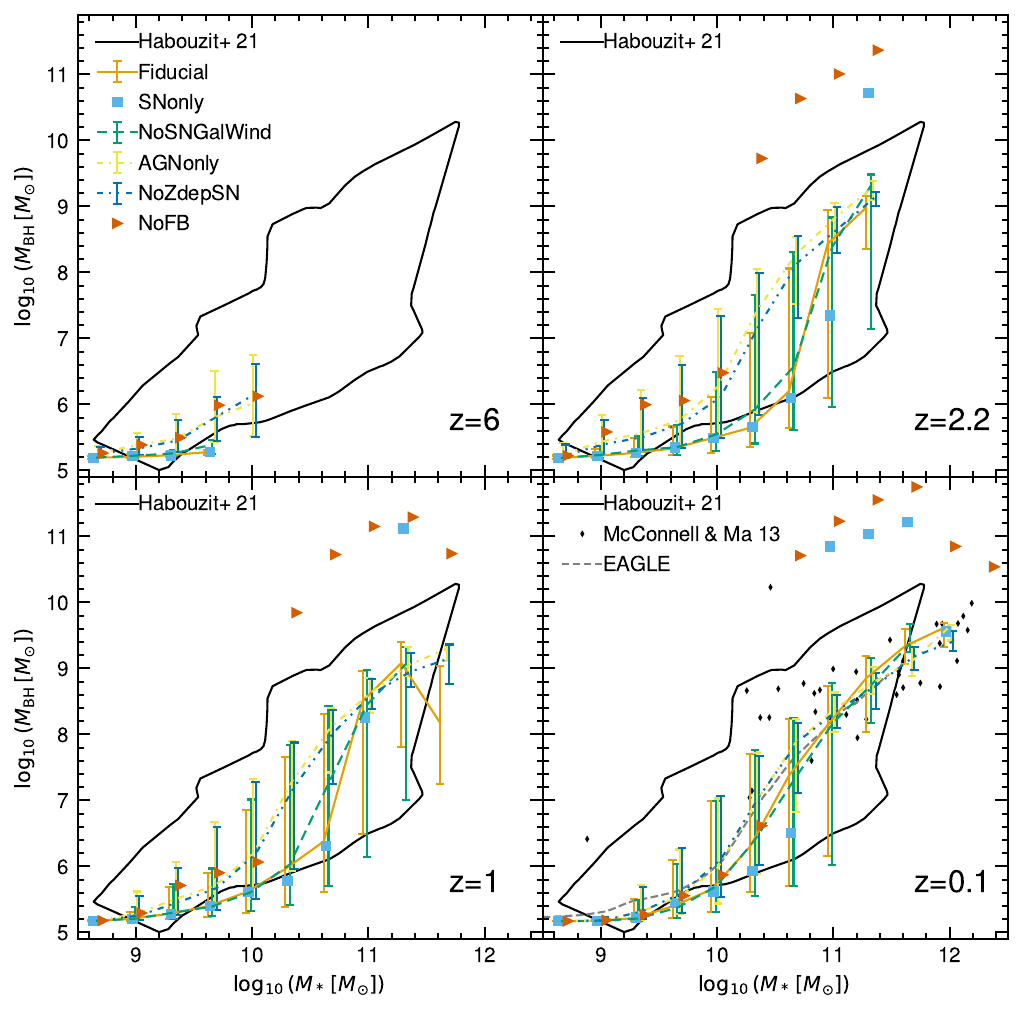}
    \caption{Redshift evolution of the BH mass--stellar mass relation. The points and error bars show the median value and 16th and 84th percentile of the most massive BH mass in galaxies in each stellar mass bin.
    For the SNonly and NoFB runs, we omit the error bars for clarity.
    The black dashed line shows the median of the EAGLE Ref run.
    The region indicated by the gray solid line covers the majority of local observational constraints compiled by \citet{2021MNRAS.503.1940H}.
    The gray diamonds are observational data by \citet{2013ApJ...764..184M}.}
    \label{fig:BH}
\end{figure*}

Figure~\ref{fig:BH} shows the BH mass as a function of the galaxy's total stellar mass.
The points and error bars show the median value and 16th and 84th percentile of the most massive BH mass in galaxies in each stellar mass bin.
The region indicated by the black solid line is the observational constraint at $z=0$ compiled by \citet{2021MNRAS.503.1940H}.
The gray diamonds are observational data by \citet{2013ApJ...764..184M}.

The runs with AGN feedback have lower BH masses at $z\lesssim3$ because the gas around BHs is blown away by the AGN feedback.
In the runs with AGN feedback, the $M_{\rm BH}$--$M_*$ relation is consistent with observations, and the medians are almost at the lower edge of the observationally constrained region at $10^9\,\Msun < M_* < 10^{10.5}\,\Msun$.
This behavior is sensitive to the choice of black hole subgrid parameters, as demonstrated in Section~\ref{sec:result_bh}.

The AGNonly run also shows higher BH mass than the Fiducial run at $z>1$, indicating that the stellar feedback inhibits the gas accretion onto BHs and suppresses BH growth.
{At $z=0.1$, the $M_{\rm BH}$--$M_*$ relation of the two runs converges. This convergence is interesting considering that the AGNonly and Fiducial run have distinct SMF (Fig.~\ref{fig:SMF}; AGNonly run significantly overpredicts the SMF relative to the Fiducial run), suggesting that the growth of supermassive BHs is self-regulated by AGN feedback, and the BH and galaxy co-evolve even if the $M_*$ increases more rapidly due to lack of SN feedback.}

\subsection{BH Parameter Variation}
\label{sec:result_bh}

\begin{figure*}
    \centering
    \includegraphics[width=0.8\textwidth]{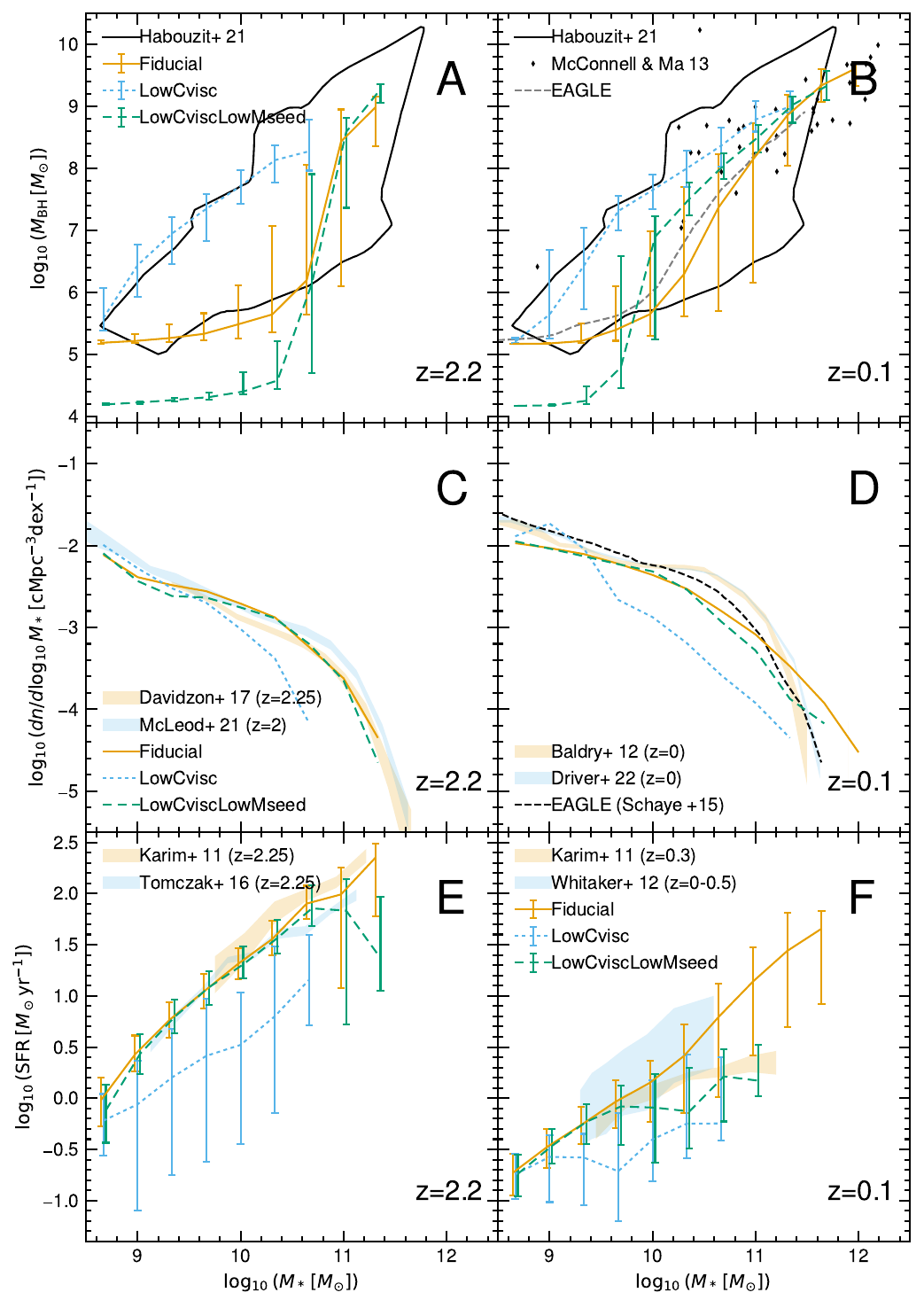}
    \caption{Comparing the $M_{\rm BH}$--$M_*$ relation, SMF, and SFMS for the runs with different BH parameters, Fiducial, LowCvisc, and LowCviscLowMseed, at $z=2.2$ and 0.1.}
    \label{fig:BHvariation}
\end{figure*}

Figure~\ref{fig:BHvariation} again shows the $M_{\rm BH}$--$M_*$ relation, SMF, and SFMS, but for the runs with different BH parameters, Fiducial, LowCvisc, and LowCviscLowMseed, at $z=2.2$ and 0.1.

Panels A and B show the $M_{\rm BH}$--$M_*$ relation.
The $C_{\rm visc}$ controls the onset of BH growth, and the BH mass starts to increase at $M_* \sim 10^9\,\Msun$ for runs with low $C_{\rm visc}$ ($=2\pi$) and at $10^{10}\,\Msun$ for runs with our default $C_{\rm visc}$ ($=200\pi$) at $z=0.1$.
The simulation data points of the Fiducial run are around the bottom of the region indicated by \citet{2021MNRAS.503.1940H}, while those of the LowCvisc run are around the top of the region.
At $M_*> 10^{11}\,\Msun$, it is interesting to see that the $M_{\rm BH}$--$M_*$ relation for all three simulations agree due to the self-regulated growth of BH.

Panels C and D show the SMF.
By comparing with panels A and B, one can see that the SMF rapidly declines at the same mass scale where the BH mass rapidly increases.
The SMF of the LowCvisc run is lower than that of other runs at $M_* > 10^{9.5}\,\Msun$ due to the early growth of the BHs and subsequent AGN feedback.
In the LowCviscLowMseed run, the BH growth is delayed due to the smaller BH seed mass, as seen in the upper panels, and it shows a similar SMF to the Fiducial run. 

Panels E and F show the impact of AGN feedback on the star-formation activity in the SFMS.
The slope of the SFMS becomes shallower at the same stellar mass where the BH begins to grow rapidly.

\subsection{Resolution Variation}

\begin{figure*}
    \centering
    \includegraphics[width=\textwidth]{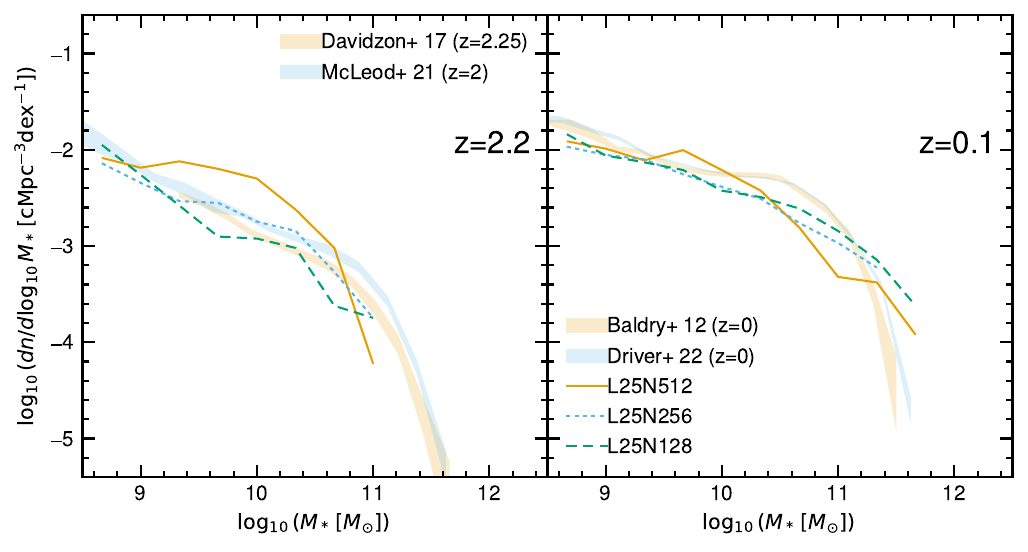}
    \caption{Galaxy stellar mass functions of the runs with different resolutions, L25N512, L25N256, and L25N128 at $z=2.2$ and 0.1.}
    \label{fig:SMFres}
\end{figure*}

Figure~\ref{fig:SMFres} shows the SMFs of the runs with different resolutions, L25N512, L25N256, and L25N128.
We used the same subgrid and numerical parameters except for the gravitational softening length in these runs.
They roughly converge and agree with the observed SMF.
The SMF slightly increases with increasing resolution at $z=2.2$.
This is attributed to the density dependence of the star formation efficiency; higher-resolution simulations can resolve higher-density star-forming clouds, which have shorter star formation timescales.

\section{Metal Enrichment in the IGM}
\label{sec:IGM}

In this section, we analyze the distribution of metals in the IGM and show the impact of SN and AGN feedback by comparing simulations with varying feedback models.

\begin{figure*}
    \begin{interactive}{animation}{fig9_anim.mp4}
    \begin{minipage}{0.85\textwidth}
        \centering        \includegraphics[width=\columnwidth]{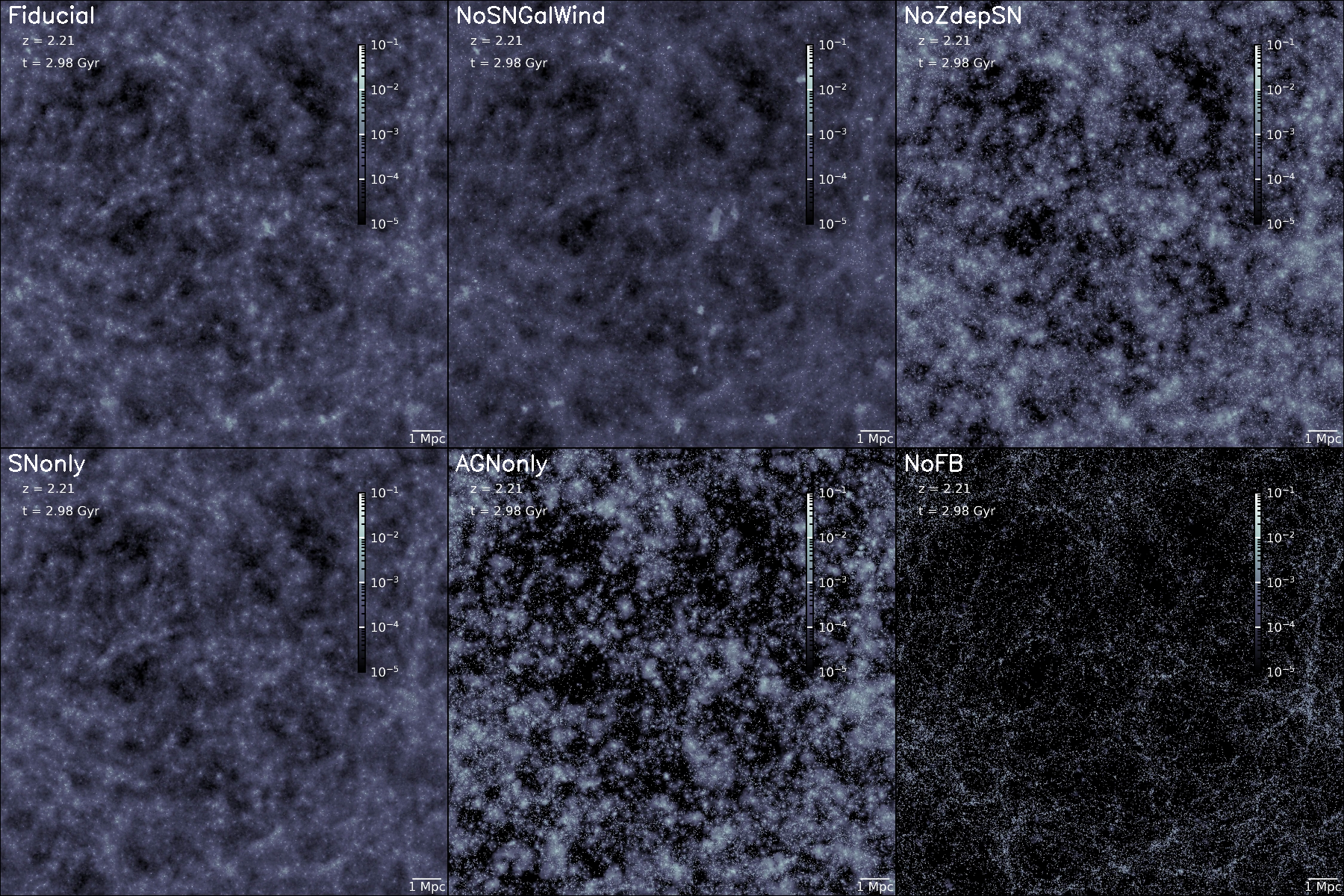}
    \end{minipage}
    \begin{minipage}{0.85\textwidth}
        \centering
        \includegraphics[width=\columnwidth]{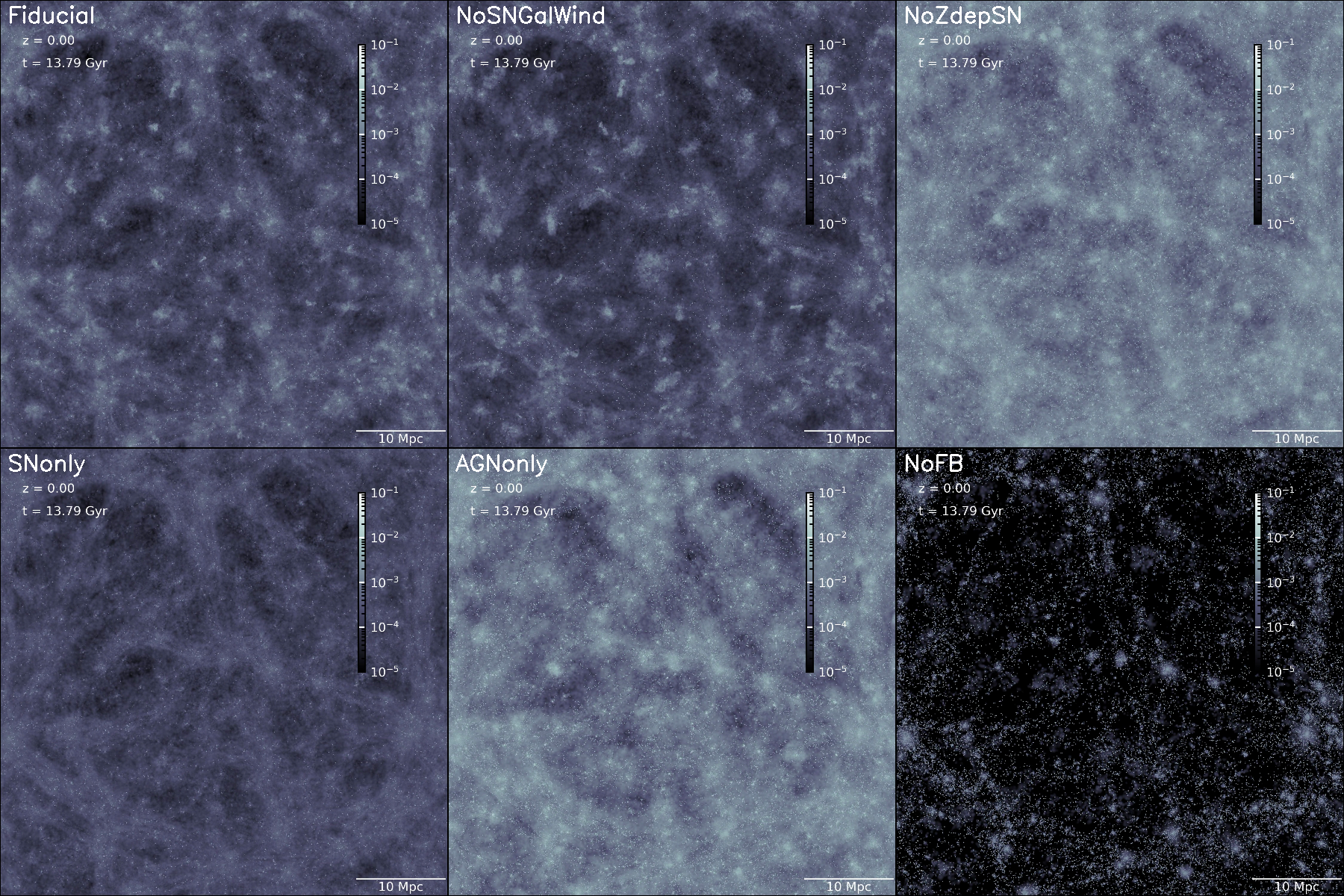}
    \end{minipage}
    \end{interactive}
    \caption{The density-weighted projection of metallicity at $z=2.2$ (top two rows) and 0 (bottom two rows). An animated version of this figure, showing the evolution of these panels from $z$ = 39 to 0, is available in the HTML version of this article.}
    \label{fig:MetalProj}
\end{figure*}

Figure~\ref{fig:MetalProj} shows the density-weighted projections of metallicity at $z=$2.2 and 0.\footnote{The movie is available at \url{https://www.yurioku.com/research}} 
{The projection depth is $50\,h^{-1}$\,Mpc.} 
Metals are distributed following the large-scale structure with visible differences due to feedback settings.

The effect of explicit modeling of SN-driven galactic wind can be seen in the comparison between the Fiducial and NoSNGalWind runs.
The SN-driven metal-rich galactic wind spread metals on $\lesssim100\,{\rm kpc}$ scale in the Fiducial run more uniformly than in the NoSNGalWind run.

Comparing the Fiducial and SNonly runs clarifies the role of AGN outflow in the chemical enrichment of the IGM.
The high-metallicity bubbles of a size of a few Mpc are formed by AGN feedback in the Fiducial run at $z=0$, but they are not visible in the SNonly run.
We modeled AGN feedback as an isotropic thermal energy injection, however the hot bubble cannot expand into the direction of the galactic disk.
Therefore, the metal-rich AGN outflow naturally expands in the bipolar direction above and below the galactic disk.

In the NoFB case, metals produced in galaxies remain there, and the distribution of metals follows that of galaxies at $z=2.2$.
It is interesting to see that metals are somewhat spread at $z=0$, which is likely due to the hydrodynamical effect in galaxy mergers, galaxy collision, and ram pressure stripping.
In other runs, feedback spread metals out to CGM and IGM.

The AGNonly {and NoZdepSN} runs show a distinctively higher metallicity than other runs at $z=0$ because star formation efficiency is higher due to the lack {or the weakness} of SN feedback, more metals are produced by stars, and AGN feedback spreads metals to IGM.
The AGN feedback begins to have an effect at $z\sim 2$, and the metal enrichment in the void region is delayed from other runs with SN feedback at $z=2.2$.

\begin{figure*}
    \centering
    \includegraphics[width=\textwidth]{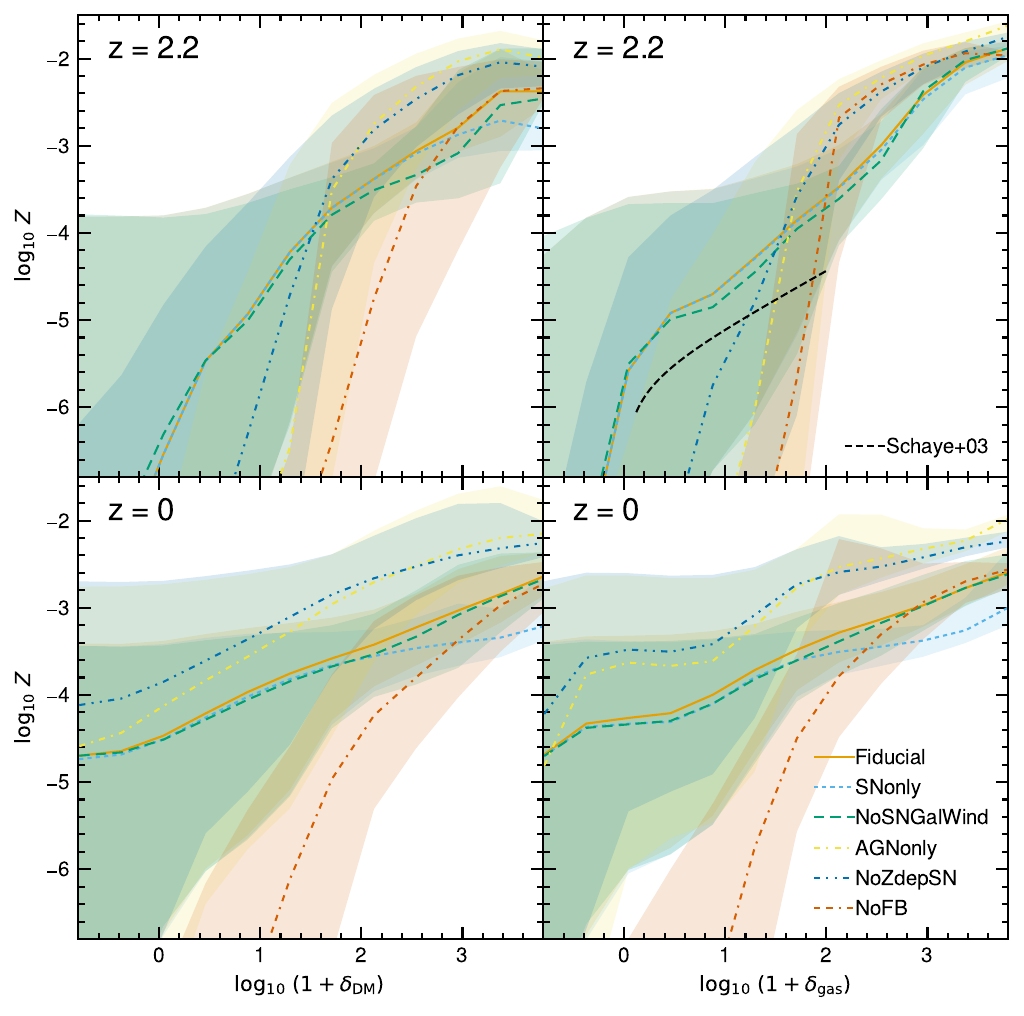}
    \caption{Metallicity as a function of dark matter (top) and gas (bottom) overdensity at $z=2.2$ and 0. The points and shades indicate the median value and 16th and 84th percentile of the metallicity.
    The black dashed line shows the observational estimate of the metallicity--overdensity relation using Ly$\alpha$ forest \citep{2003ApJ...596..768S}.}
    \label{fig:DeltaZ}
\end{figure*}

Figure~\ref{fig:DeltaZ} shows the IGM metallicity as a function of the dark matter and gas overdensity.
The black dashed line shows the best-fit line for the observational estimate of the metallicity--overdensity relation from the carbon abundance in the Ly$\alpha$ forest \citep{2003ApJ...596..768S}, ${\rm [C/H]} = -3.47 + 0.08(z-3) + 0.65(\log \delta - 0.5)$, where $\delta$ is the baryon overdensity and $z$ is the redshift.
We generate a uniform cartesian mesh with $512^3$ voxels and compute dark matter density, gas density, and gas metallicity as described in Section~\ref{sec:method_analysis}.

{The Fiducial, NoSNGalWind, and SNonly runs show very similar results with each other, with higher metallicities at low-$\delta$ than other runs at $z=2.2$. 
These runs suggest that thermal galactic wind and AGN feedback play minor roles in the metal enrichment of the IGM, and the SN mechanical feedback is the dominant mechanism to enrich the IGM among other feedback processes at $z>2$.
The slope of these runs shows a good agreement with the observation at $z=2.2$, which is encouraging. 
The AGN feedback effect in enrichment is prominent at $z=0$ at high-$\delta$ end, where the median metallicity in the SNonly run is lower by 0.5 dex than the Fiducial run.
Our simulations suggest that the low-overdensity region is mostly enriched by SN mechanical feedback from low-mass galaxies, but this statement needs to be backed by further analysis of galactic wind properties and metal--galaxy cross-correlation.}

{Compared to the Fidicial run, the AGNonly run shows higher metallicity at $\log_{10}(1+\delta)>1.5$ at $z=2.2$, because more metals are produced and confined in galaxies due to the absence of stellar feedback. The AGNs are not active enough to enrich the IGM at $z=2.2$, and the metallicity drops at lower $\delta$.}
Below $z\sim2$, the AGN feedback kicks in, spreads metals to IGM, and increases metallicity at low overdensity regions, which cannot be seen in the NoFB run.
{The NoZdepSN run shows a similar metallicity--overdensity relation to the AGNonly run but has higher metallicity at low overdensity.
Although the SN feedback in the NoZdepSN run is weak to suppress star formation at high redshift, it powers galactic winds and enriches the IGM further than in the AGNonly run.
}

\begin{figure*}
    \centering
    \includegraphics[width=\textwidth]{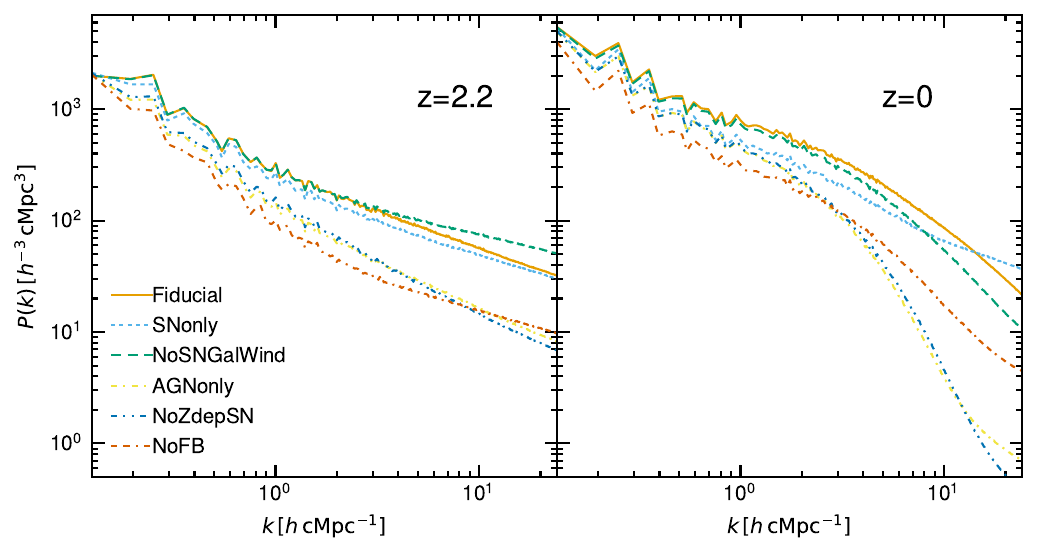}
    \caption{Power spectra of the metal density field at $z=2.2$ (left) and 0 (right). The upper limit of the abscissa (wavenumber) is cut off at $k=k_{\rm Nyq}/4$, where $k_{\rm Nyq}$ is the Nyquist frequency.}
    \label{fig:PowerMetal}
\end{figure*}

Figure \ref{fig:PowerMetal} shows the power spectra of the metal density field at $z=2.2$ and 0, which allows us to examine the different effects of feedback models statistically.
We use the cosmological analysis toolkit \textsc{nbodykit}\footnote{The code's website is \url{https://nbodykit.readthedocs.io/en/latest/index.html}} to compute the power spectra on $1536^3$ mesh, and the upper limit of the abscissa (wavenumber) is cut off at $k=k_{\rm Nyq}/4$, where $k_{\rm Nyq}=\pi N_{\rm mesh} / L_{\rm box}$ is the Nyquist frequency.

At $z=2.2$, the Fiducial run has lower power at $k>3\,h\,{\rm cMpc}^{-1}$ compared to the NoSNGalWind run, indicating that the galactic wind transport metals and smooth out the metal distribution in CGM.
The effect of AGN can be seen around $k\sim3\,h\,{\rm cMpc}^{-1}$ as a slight enhancement of the power in the Fiducial run relative to the SNonly run due to the AGN-driven outflows entraining metals out to a few Mpc scale.

At $z=2.2$, metals are more diffused in the Fiducial run compared to the NoSNGalWind run, hence lower power on small scales.
{In the NoSNGalWind run, more metals are retained inside dark matter halo at $z \sim 2$, and those metals are transported into the IGM by AGN feedback by $z=0$, leading to a lower power on small scales compared to the Fiducial run.}

{The amplitude of the power spectrum of the NoFB run is lower than the Fiducial run, which is somewhat counterintuitive since the metals are highly concentrated in galaxies in the NoFB run. 
This could be because the metals are locked up in stars and BHs. Indeed, the total metal mass in the gas phase of the whole simulation box in the NoFB run is smaller than the Fiducial run by a factor of 0.78 at $z=0$, whereas the total cumulative metal mass produced by stars is larger by a factor of 3.2.}

\section{Discussion and Summary}
\label{sec:summary}

In this paper, we introduce a new suite of cosmological hydrodynamical simulations {\osakasim}, powered by the \textsc{GADGET4-Osaka} code.
Building on the supernova feedback model detailed in \citet{2022ApJS..262....9O}, we have incorporated the galactic wind model from the \textsc{TIGRESS} simulations \citep{2020ApJ...903L..34K}.
Our simulations also feature an updated chemical evolution model, which integrates a metallicity- and redshift-dependent IMF based on the star cluster formation simulation by \citet{2022MNRAS.514.4639C}, along with a metallicity-dependent HN fraction.  This refinement significantly enhances the specific SN energy of low-metallicity stars by more than an order of magnitude (Fig.~\ref{fig:imf}).
Furthermore, we implement an AGN feedback model,  following the methodologies outlined in \citet{2015MNRAS.454.1038R, 2015MNRAS.446..521S, 2015MNRAS.450.1937C}.  This model approaches the AGN feedback as a stochastic thermal energy injection process.

The main points of this paper can be summarized as follows:
\begin{itemize}
    \item Our simulations reproduce the observed SMF across a broad spectrum of galaxy stellar masses, ranging from $10^8\,\Msun$ to $10^{11}\,\Msun$ as illustrated in Figure~\ref{fig:SMF}. Notably, at $z\lesssim3$, our SN feedback model, incorporating a top-heavy IMF, effectively suppresses the formation of low-mass galaxies at $M_*<10^{10.5}\,\Msun$. At $z\sim0.1$, AGN feedback plays a crucial role in reducing the number of massive galaxies with $M_*\gtrsim10^{11.8}\,\Msun$ (bottom right panel of Figure~\ref{fig:SMF}).
    We anticipate that the influence of AGN feedback will be even more pronounced at the higher-mass end of the SMF in a larger cosmological box of $100\,{\rm Mpc}$, which we will perform in the near future.
    
    {The top-heavy IMF and high HN fraction are suggested by both simulations \citep{2006ApJ...653.1145K, 2022MNRAS.514.4639C, 2024MNRAS.530.2453C, 2023MNRAS.524.2594K,2023MNRAS.525.4832Y} and recent \textit{JWST} observations \citep{2023arXiv231104279F, 2023MNRAS.518.6011D, 2023MNRAS.519.3064F, 2023ApJS..265....5H, 2024ApJ...960...56H,2022ApJ...925..111I, 2024ApJ...962...50W, 2024arXiv240714470N}. 
    However, we note that the necessity of the top-heavy IMF and high HN fraction at low metallicities is not a robust conclusion of our work; the SMF is reproduced by a combination of our subgrid physics. It is possible that another feedback model, star formation recipe, or black hole seeding and accretion model can reproduce the SMF without adopting the top-heavy IMF.}
    
    \item  At $z\sim6$, our SN feedback model appears to be a bit too strong, leading to an underprediction of galaxies with $M_*<10^{9.5}\,\Msun$ despite the good agreement in the SFMS (top left panel of Figure~\ref{fig:SFMS}).
    Interestingly, runs without SN mechanical feedback demonstrate a closer agreement with observational data. \\
    Moreover, recent observations by the \textit{James Webb Space Telescope} (\textit{JWST}) have unveiled  more numerous massive galaxies at $z>7$ than expected from cosmological simulations \citep{2023ApJS..265....5H, 2023MNRAS.523.1009B, 2023arXiv231104279F, 2023arXiv231205030C, 2024MNRAS.527.5929Y}.
    At $z\gtrsim6$, the discrepancy between our simulation results and observational data becomes more pronounced. 
    \citet{2023MNRAS.523.3201D} have introduced the concept of a `feedback-free starburst' scenario.  This theory suggests that gas in high-$z$ halo can be efficiently converted into stars with minimal feedback effects, primarily due to the conditions of high density and low metallicity prevalent at these early cosmic times. 
    The closer agreement of the NoFB run with observations lends support to this feedback-free starburst hypothesis. However, it remains uncertain whether this scenario can consistently account for the observed SMF at lower redshifts.  This uncertainty points to a potentially critical area of investigation in understanding the evolution of galaxies across different epochs.

    \item Our simulations generally agree with the observed SFMS across all redshifts, except for the runs without SN feedback (AGNonly and NoFB runs) as depicted in Figure~\ref{fig:SFMS}.
    The latter two runs significantly overestimate the SMF and MZR, while simultaneously underestimating the SFMS and IGM metallicity. This discrepancy underscores the critical role of SN feedback in cosmological galaxy evolution.
    
    \item Our SMBH result at $z=0.1$ shows a general agreement with the EAGLE Ref run within the mass range of $M_*=10^{8} - 10^{12}\,\Msun$ and $M_{\rm BH}=10^{5} - 10^{10}\,\Msun$ (Fig.~\ref{fig:BH}).
    With a low $C_{\rm visc}$ parameter, the mass accretion onto BH in the LowCvisc run is enhanced relative to the Fiducial run, leading to an overestimation of $M_{\rm BH}$ and a corresponding underestimation of both the SMF and SFMS (Fig.~\ref{fig:BHvariation}).
    Furthermore, the runs without AGN feedback (SNonly and NoFB runs) consistently overpredict the $M_{\rm BH}$.

    \item We find that the SN feedback is the dominant mechanism to transport metals into the IGM (Fig.~\ref{fig:DeltaZ}).
    In addition, the AGN feedback generates bipolar, metal-rich outflow (Fiducial and NoSNGalWind panels of Fig.~\ref{fig:MetalProj}), effectively enriching the CGM and extending into the IGM up to a few Mpc scales, as observed in the power spectrum of metal distribution with scales $k<10\,h\,{\rm Mpc}^{-1}$ (right panel of Fig.~\ref{fig:PowerMetal}).
\end{itemize}

In this paper, we primarily focused on the broad distributions of metals as functions of overdensities and spatial scales.
Given that our simulation tracks multiple chemical species, we are well positioned to extend our analysis to more detailed ionization calculations for individual elements. This will enable us to conduct comparative studies with a variety of metal absorption lines observed in quasar spectra.
Additionally, our simulation incorporates sophisticated models for the formation and destruction of dust \citep{2017MNRAS.466..105A, 2022MNRAS.514.1441R}, which allows us to investigate the dust distribution within galaxies, CGM, and IGM in the future.
By conducting cross-correlation studies between galaxies, H\,\textsc{i}, dust, and metals, we anticipate gaining a more comprehensive understanding of galaxy evolution and the dynamics of the baryonic universe.

\section*{Data Availability}

{The scripts and data used to generate the figures in this article are available on GitHub\footnote{Supplemental materials: 
\url{https://github.com/YuriOku/figure_introducing-crocodile/tree/main}}.
The simulation data will be available upon reasonable request to the authors through the simulation homepage\footnote{Simulation homepage: \url{https://sites.google.com/view/crocodilesimulation/home}}.}

\section*{Acknowledgements}
{We thank the anonymous referee for helpful comments that improved the manuscript significantly.}
Our numerical simulations and analyses were carried out on the XC50 systems at the Center for Computational Astrophysics (CfCA) of the National Astronomical Observatory of Japan (NAOJ) and {\sc SQUID} at the Cybermedia Center, Osaka University as part of the HPCI System Research Project (hp220044, hp230089, hp240141).
This work is supported by the JSPS KAKENHI Grant Number 21J20930, 22KJ2072 (Y.O.), 19H05810, 20H00180, 22K21349, 24H00002, 24H00241 (K.N.). 
K.N. acknowledges the support from the Kavli IPMU, World Premier Research Center Initiative (WPI), UTIAS, the University of Tokyo.  

\appendix
\section{Hydrodynamics}
\label{sec:hydrodynamics}
Here, we describe the numerical methodology of the cosmological hydrodynamic simulation by \textsc{GADGET4-Osaka}.
\subsection{SPH}
We determine the smoothing length by
\begin{equation}
    \frac{4}{3}\pi h_i^3 n_i = N_{\rm ngb},
\end{equation}
where 
\begin{equation}
    n_i = \sum_{j=1}^N W_{ij}(h_i)
\end{equation}
is the SPH particle number density, and $W_{ij}(h_i) \equiv W(|\bm{r}_i - \bm{r}_j|; h_i)$ is the kernel function. We use the Wendland C4 kernel function \citep{Wendland1995PiecewisePP, 2012MNRAS.425.1068D} with neighbor number $N_{\rm ngb} = 120\pm2$.

We use the pressure-energy formulation of SPH \citep{2013MNRAS.428.2840H, 2013ApJ...768...44S}.
The equation of motion of SPH particle is
\begin{equation}
\label{eq:sph_eom}
\begin{split}
    \frac{dv_i}{dt} = -\sum_{j=1}^N m_j \left[f_{ij}\frac{u_j}{u_i}\frac{\bar{P}_i}{\bar{\rho}_i^2}\nabla_i W_{ij}(h_i)\right.\\
    \left.+ f_{ji}\frac{u_i}{u_j}\frac{\bar{P}_j}{\bar{\rho}_j^2}\nabla_i W_{ij}(h_j) \right].
\end{split}
\end{equation}
The energy equation is
\begin{equation}
\label{eq:sph_energyeq}
\frac{du_i}{dt} = \sum_{j=1}^N m_j f_{ij}\frac{u_j}{u_i}\frac{\bar{P}_i}{\bar{\rho}_i^2}(\bm{v}_i - \bm{v}_j) \nabla_i W_{ij}(h_i), 
\end{equation}
where 
\begin{equation}
    \bar{P}_i= \sum_{j=1}^N (\gamma - 1) m_j u_j W_{ij}(h_i)
    \label{eq:smthpress}
\end{equation}
is the smoothed pressure, $\bar{\rho}_i = \bar{P}_i/[(\gamma - 1) u_i]$ is the internal-energy-weighted smoothed density, and
\begin{equation}
    f_{ij} = \left[\frac{h_i}{3(\gamma - 1)n_i m_j u_i}\frac{\partial \bar{P}_i}{\partial h_i} \right] \left(1 + \frac{h_i}{3n_i}\frac{\partial n_i}{\partial h_i}\right)^{-1}
\end{equation}
is the grad-$h$ term, which accounts for the variation of the smoothing length.\footnote{Note that entropy and energy are simultaneously conserved in the pressure-energy SPH formulation derived from Lagrangian \citep{2013MNRAS.428.2840H}.}

We did not use the pressure-entropy formulation, another pressure-based SPH variant implemented in original \textsc{GADGET-4}.
The pressure-entropy SPH cannot correctly increase or decrease energy without a costly iterative treatment when energy is added or reduced by subgrid physics \citep{2021MNRAS.505.2316B}.
This is because the smoothed density $\bar{\rho}_i = (1/A_i^{1/\gamma}) \sum_j m_j A_j^{1/\gamma} W_{ij}(h_i)$ is coupled to the entropy $A_i$, and one cannot change the particle's energy $u_i = A_i \bar{\rho}_i^\gamma$ as desired by simply updating entropy.
In the pressure-energy formulation, each SPH particle explicitly has an internal energy variable, and modifying the energy is straightforward.

Artificial viscosity is computed with velocities linearly reconstructed at the particle mid-point \citep{2017JCoPh.332..160F, 2020MNRAS.498.4230R}.
The high-order estimator of the velocity gradient by \citet{2014MNRAS.443.1173H} is used to reconstruct the velocity field.
Artificial conduction \citep{2008JCoPh.22710040P} is included with the velocity-based conductivity signal speed \citep{2008MNRAS.387..427W} and the conduction limiter by \citet{2022MNRAS.511.2367B}.

The density and hydro force evaluation kernel is vectorized using the AVX-512 vector extension instruction set, which can perform arithmetic or logical instruction for eight 64-bit double variables in a single instruction.
The SPH calculation consists of a doubly-nested loop; 1) for all active particles, 2) we take kernel-weighted sums.
\citet{2021MNRAS.506.2871S} reported that the original \textsc{GADGET-4} code shows no significant improvement in the computational speed using the AVX2 instruction set.
This would be likely because they vectorized the innermost SPH loop, where the memory bandwidth limits the calculation speed.
Instead, we vectorize the outer loop by simultaneously processing a group of SPH particles \citep{1990JCoPh..87..161B}.

At the beginning of every SPH evaluation loop, we form groups consisting of, at most, eight particles using the indices of SPH particles.
The indices are assigned using the Peano-Hilbert curve when domain decomposition is performed, and the particles that are close on the index list are also close in positions \citep{2005MNRAS.361..776S}.
To form groups, we check the particles in the order of indices and add a particle to a group if their regions to search for neighbors overlap each other.

\subsection{Jeans pressure floor}
\label{sec:jeans_floor}
The non-thermal Jeans pressure floor is introduced to avoid numerical fragmentation. This additional pressure term can be interpreted as unresolved interstellar turbulence and allows us to follow the thermal evolution of atomic gas to low temperature.

The floor is usually implemented as 
\begin{equation}
    P_{\rm hydro} = \max\{P, P_{\rm Jeans}\},
\end{equation}
and
\begin{equation}
   P_{\rm Jeans} = (\gamma \pi)^{-1} G N_{\rm Jeans}^2 \rho^2 \Delta x^2,
\end{equation}
where $\gamma=5/3$ is the adiabatic index, $N_{\rm Jeans}=4$ is the Jeans number \citep{1997ApJ...489L.179T}, and $\Delta x = \max(\epsilon_{\rm grav}, (m/\rho)^{1/3}$) is the local resolution, which is the larger of the gravitational softening length and the mean particle separation.
The $P$ is the particle's pressure, and $P_{\rm hydro}$ is the pressure used in the hydro force evaluation.
This implementation works as intended in the density-based SPH formulation; however, care is necessary when used with the pressure-based formulation.

The pressure used in the pressure-based SPH (Eq.\,\ref{eq:smthpress}) is obtained from the kernel sum of internal energy.
Modifying the pressure alone causes inconsistency between the internal energy and pressure. 
The appropriate implementation is to introduce a floor in the internal energy.\footnote{In the density-based SPH, modifying the particle's pressure is equivalent to modifying its internal energy.}
For each SPH particle, we store floored internal energy, $u_{\rm hydro} = \max\{u, P_{\rm Jeans}/[(\gamma - 1)\rho]\}$, where $\rho = \sum_j^N m_j W_{ij}(h_i)$ is physical density, in addition to the original internal energy. 
We use the floored internal energy to compute the smoothed pressure and hydro force.

\subsection{Time Integration}

In \textsc{GADGET-4}, the hydrodynamical force is integrated using the second-order predictor-corrector method \citep{2021MNRAS.506.2871S}.
In the following, we call the integration width on the timeline a time step, and the point on the timeline where a previous time step ends and the next step begins a synchronization point.

We use the time step limiter that wakes up inactive SPH particles to force a time step constraint \citep{2009ApJ...697L..99S, 2012MNRAS.419..465D}. 
For the constraint, we use the signal velocity timestep limiter \citep{2010MNRAS.401..791S,2021MNRAS.506.2871S} instead of constraining the neighbor's time step within a fixed factor of difference. 
The signal velocity limiter is more general and flexible than the fixed factor limiter, as demonstrated in \citet{2010MNRAS.401..791S}.
The implementation of the signal velocity time step limiter in the original \textsc{GADGET-4} is one-sided, and information on neighboring particles is used to constrain the new time step of a particle. 
We extended the limiter to a mutual constraint in our \textsc{GADGET4-Osaka}; when the particle $i$ is updated, we search its neighbor $j$ and impose the time step constraint $\Delta t_j <  C_{\rm CFL} \Delta t_j^{\rm signal}$, where
\begin{equation}
    \Delta t_j^{\rm signal} = \frac{2h_j + r_{ij}}{c_i^{\rm snd} + c_j^{\rm snd} - \bm{r}_{ij}\cdot\bm{v}_{ij}/r_{ij}},
    \label{eq:dt_signal}
\end{equation}
and $\bm{r}_{ij}=\bm{r}_{i}-\bm{r}_{j}$, $\bm{v}_{ij}=\bm{v}_{i}-\bm{v}_{j}$, $r_{ij}=|\bm{r}_{ij}|$, and $c^{\rm snd}_i$ is the sound speed of $i$-th particle.
This wake-up scheme ensures that all SPH particles maintain the signal velocity time step criterion even when the velocity and the sound speed of neighboring particles are updated due to feedback.

The evaluation of $\Delta t^{\rm signal}_j$ is carried out by walking on the oct-tree used for the neighbor search of SPH particles.
For each node of the oct-tree, we store the maximum signal time step $\Delta t^{\rm signal}_{\rm max}$, the maximum of $d^{\rm signal} = c^{\rm snd} \Delta t^{\rm signal}-2h$, the maximum sound speed $c^{\rm snd}_{\rm max}$, and the maximum and the minimum of velocity in x-, y-, and z-direction $(v_{\rm max}^x, v_{\rm min}^x, v_{\rm max}^y, v_{\rm min}^y, v_{\rm max}^z, v_{\rm min}^z)$ of SPH particles in the node.
When we encounter a node in the tree walk, we compute the node opening criterion,
\begin{equation}
    d_{\rm min} < d^{\rm signal}_{\rm max} + \Delta t^{\rm signal}_{\rm max}(c^{\rm snd}_i - v^{\rm rel}_{\rm min}),
    \label{eq:dmin_j}
\end{equation}
where $v^{\rm rel}_{\rm min} = \min_j(\bm{r}_{ij}\cdot\bm{v}_{ij}/r_{ij})$ is the minimum relative velocity, i.e., the largest approaching velocity, between particle $i$ and particles in the node, which can be constrained using the maximum and minimum velocity of particles in the node.
The $d_{\rm min}$ is the smallest distance from the particle $i$ to the node.
If the opening condition is fulfilled, we open the node and continue walking on its daughter nodes.
When we encounter a particle $j$, we compute Eq.\,(\ref{eq:dt_signal}) and update $\Delta t^{\rm signal}_j$ if it is smaller.
If $C_{\rm CFL} \Delta t^{\rm signal}_j$ is smaller than particle $j$'s time step $\Delta t_j$, we flag the particle.
At the beginning of every synchronization point, we check flagged particles and wake them up if it is the point where a particle with the time step of $C_{\rm CFL} \Delta t^{\rm signal}_j$ should be synchronized.

We note that the tree walk is carried out along with the evaluation of $\Delta t^{\rm signal}_i$, and its node opening criterion is
\begin{equation}
    d_{\rm min} < d^{\rm signal}_{\rm i} + \Delta t^{\rm signal}_i(c^{\rm snd}_{\rm max} - v^{\rm rel}_{\rm min}).
    \label{eq:dmin_i}
\end{equation}
We actually open the tree node if the criteria (\ref{eq:dmin_j}) or (\ref{eq:dmin_i}) are fulfilled, and when we encounter a particle, we compute $\Delta t^{\rm signal}_i$ and update if it is smaller.

We also note that the number of neighbor particles can be large when there are particles with high temperatures and/or high relative velocities.
In \textsc{GADGET-4}, the default tree walk method is to build a local essential tree, which requires a memory buffer to allocate data of tree nodes and particles on other memory space.
When we have a large number of neighbors, the construction of the local essential tree can fail due to memory shortage.
We thus implemented the tree-based time step limiter using the generic communication pattern, a C++ class of MPI communication routine available in \textsc{GADGET-4}.
The tree walk can be costly when we have many neighbors, but the signal velocity time step limiter chooses the optimal time step for each particle, and unnecessary calculation can be avoided to reduce the computational cost.

In addition to the signal velocity time step constraint, we activate particles that can be subject to stellar or AGN feedback.
At the beginning of a synchronization point, we drift neighbor particles around the feedback site and find particles that will receive feedback energy.
The neighbor particles are woken up if they are inactive.
This ensures that the feedback effect is reflected in hydrodynamics without delay.

\bibliography{ref}
\bibliographystyle{aasjournal}

\end{document}